\documentclass[aps, prd, onecolumn, tightenlines, notitlepage, superscriptaddress, nofootinbib, preprintnumbers, floatfix,showkeys,11pt]{revtex4-2}

\usepackage[normalem]{ulem}
\usepackage{amstext}
\usepackage{amssymb}
\usepackage{amsmath}
\usepackage{graphicx}
\graphicspath{{plots/}}
\usepackage{url}
\usepackage{color}
\usepackage{ulem}
\usepackage[utf8]{inputenc}
\pdfoutput=1
\usepackage{textcomp}

\usepackage{epsfig,amsfonts,mathrsfs,amsmath,amssymb,graphicx,color,slashed,multirow}
\usepackage{amsmath,latexsym,amssymb,graphicx,slashed,color,enumerate,url,cancel,gensymb}
\usepackage{textcomp}

\usepackage[x11names]{xcolor}
\usepackage[colorlinks]{hyperref}

\usepackage{textgreek} 
\usepackage{caption, subcaption} 
\usepackage{booktabs} 

\hypersetup{colorlinks,citecolor= nicered,linkcolor= blue}
\definecolor{nicered}{rgb}{0.7,0.1,0.1}
\definecolor{nicegreen}{rgb}{0.1,0.5,0.1}

\def\cevns{CE$\nu$NS}

\def\d{\mathrm{d}}

\definecolor{blue(ncs)}{rgb}{0.0, 0.53, 0.74}
\definecolor{chromeyellow}{rgb}{1.0, 0.56, 0.0}
\definecolor{amber(sae/ece)}{rgb}{1.0, 0.49, 0.0}

\newcommand{\qtransfer}{\left|\mathbf{q}\right|}

  \AtBeginDocument{\hypersetup{citecolor=chromeyellow,linkcolor=chromeyellow,urlcolor=chromeyellow}}
\usepackage{appendix}

\begin{document}

\title{{\LARGE COHERENT production of a dark fermion}}
\author{Pablo M. Candela}\email{pamuca@ific.uv.es}
\affiliation{Instituto de F\'{i}sica Corpuscular (CSIC-Universitat de Val\`{e}ncia), Parc Cient\'ific UV C/ Catedr\'atico Jos\'e Beltr\'an, 2 E-46980 Paterna (Valencia) - Spain}
\author{Valentina De Romeri}\email{deromeri@ific.uv.es}
\affiliation{Instituto de F\'{i}sica Corpuscular (CSIC-Universitat de Val\`{e}ncia), Parc Cient\'ific UV C/ Catedr\'atico Jos\'e Beltr\'an, 2 E-46980 Paterna (Valencia) - Spain}
\author{Dimitrios K. Papoulias}\email{dkpapoulias@phys.uoa.gr}
\affiliation{Department of Physics, National and Kapodistrian University
of Athens, Zografou Campus GR-15772 Athens, Greece}


\begin{abstract}
We consider the possible production of a new MeV-scale fermion at the COHERENT experiment. The new fermion, belonging to a dark sector, can be produced through the up-scattering process of neutrinos off the nuclei and the electrons of the detector material, via the exchange of a light vector or scalar mediator. We perform a detailed statistical analysis of the combined COHERENT CsI and LAr datasets and obtain up-to-date constraints on the couplings and masses of the dark fermion and mediators. We finally briefly comment about the stability of the dark fermion.
\end{abstract}
\maketitle

\section{Introduction}

The neutrino physics picture, while not yet complete, is continuously improving thanks to multiple experimental efforts. On the one hand, there are oscillation experiments~\cite{Kajita:2016cak,McDonald:2016ixn} which aim at a precise determination of all neutrino mixing parameters~\cite{deSalas:2020pgw,Esteban:2020cvm}. On the other hand, experiments exploiting neutral current neutrino interactions have been proven capable of providing valuable and complementary information to this picture~\cite{SajjadAthar:2022pjt}. Among these, the recent observation of
coherent elastic neutrino-nucleus scattering (\cevns), about four decades after its first theoretical prediction~\cite{Freedman:1973yd}, has opened the window to a plethora of physics opportunities~\cite{Abdullah:2022zue, Papoulias:2019xaw}. 
The \cevns~process has been observed by the COHERENT experiment using two different detectors, one made out of cesium iodide (CsI)~\cite{COHERENT:2017ipa,COHERENT:2021xmm} and the other of liquid argon (LAr)~\cite{COHERENT:2020iec}. The COHERENT experiment is one of the few that currently use neutrinos from a $\pi$-DAR (pion decay-at-rest) source. In this particular case, neutrinos are produced at the Spallation Neutron Source (SNS). Other \cevns~experiments exploiting $\pi$-DAR neutrinos are the COHERENT CAPTAIN-Mills (CCM) at the LANSCE Lujan Center~\cite{CCM:2021leg} and the planned European Spallation Source~\cite{Baxter:2019mcx}. Several other experiments currently underway also aim at observing the \cevns~process. They all rely on a different neutrino source, namely nuclear reactors, and include facilities like CONNIE~\cite{CONNIE:2021ggh}, CONUS~\cite{CONUS:2021dwh}, $\nu$GEN~\cite{nuGeN:2022bmg},
MINER~\cite{MINER:2016igy}, RICOCHET~\cite{Billard:2016giu}, NUCLEUS~\cite{Strauss:2017cuu}, TEXONO~\cite{Wong:2015kgl}, vIOLETA~\cite{Fernandez-Moroni:2020yyl}, RED-100~\cite{Akimov:2022xvr}, NEON~\cite{NEON:2022hbk}, NEWS-G~\cite{NEWS-G:2021mhf} and
the Scintillating Bubble Chamber (SBC)~\cite{SBC:2021yal}. Let us highlight a recent result reported by the Dresden-II Collaboration, claiming a very strong preference for the presence of a \cevns~component in their data using reactor antineutrinos~\cite{Colaresi:2022obx}. The low-energy tail of more energetic neutrino sources can also be exploited to study \cevns. This is the case, for instance, of the $\nu$BDX-DRIFT Collaboration which aims at observing \cevns~ induced by decay-in-flight neutrinos produced at the Long Baseline Neutrino Facility (LBNF), using a directional time projection chamber~\cite{AristizabalSierra:2021uob, AristizabalSierra:2022jgg}.

Measurements of \cevns~provide many opportunities for precision tests of the Standard Model (SM) parameters~\cite{Papoulias:2019lfi,Coloma:2020nhf,Cadeddu:2021ijh,Majumdar:2022nby,AristizabalSierra:2022axl,DeRomeri:2022twg,Sierra:2023pnf,AtzoriCorona:2023ktl}, but also for probing the existence of new physics, for example in the form of new interactions~\cite{Barranco:2005yy,Ohlsson:2012kf,Miranda:2015dra,Dent:2016wcr,Farzan:2017xzy,Liao:2017uzy,AristizabalSierra:2018eqm,Abdullah:2018ykz,Billard:2018jnl,Denton:2018xmq,Han:2019zkz,Abdullah:2018ykz,Giunti:2019xpr,AristizabalSierra:2019ykk,AristizabalSierra:2019ufd,Denton:2020hop,Flores:2020lji,Cadeddu:2020nbr,Amaral:2021rzw,Flores:2021kzl,AtzoriCorona:2022moj,DeRomeri:2022twg,Elpe:2022hqp,Breso-Pla:2023tnz}, nontrivial electromagnetic properties~\cite{Schechter:1981hw,Pal:1981rm,Kayser:1982br,Nieves:1981zt,Shrock:1982sc,Kosmas:2015sqa,Canas:2015yoa,Miranda:2019wdy,Miranda:2020kwy} or new states~\cite{Kosmas:2017zbh,Brdar:2018qqj,Blanco:2019vyp,Miranda:2020syh,Miranda:2021kre,Bolton:2021pey,Chao:2021bvq,Chen:2021uuw,AristizabalSierra:2021fuc,DeRomeri:2022twg,Calabrese:2022mnp}. Interestingly, new states with masses lying at the MeV scale (dictated by kinematics arguments) can be produced at accelerator-based facilities like COHERENT. One example is the possible production of a heavy neutral lepton due to the presence of active-sterile neutrino transition magnetic moments, via up-scattering processes~\cite{McKeen:2010rx,Bolton:2021pey,DeRomeri:2022twg}. In full generality, not only sterile neutrinos but also other new particles belonging to a dark sector can be within the reach of \cevns~experiments. In some cases, they can constitute all or part of the dark matter (DM) of the Universe. Indeed, sub-GeV DM particles, while too light to be observed in many conventional DM detectors, may be produced at accelerators~\cite{deNiverville:2011it,Berlin:2020uwy,Batell:2022xau}. Various searches for sub-GeV DM particles have been performed both at COHERENT and at CCM~\cite{Ge:2017mcq,PhysRevD.106.012001,COHERENT:2019kwz,COHERENT:2021pvd,CCM:2021yzc}, leveraging the characteristic timing composition of $\pi$-DAR sources which allow to reduce systematic uncertainties~\cite{Dutta:2019nbn,Dutta:2020vop}. Among low-mass DM scenarios, a dark photon and a leptophobic portal have been previously investigated~\cite{deNiverville:2015mwa,COHERENT:2022pli,PhysRevLett.124.121802,CCM:2021yzc,Dutta:2023fij}.

In this paper, we go beyond the photon- or dark photon-mediated production of new dark-sector particles, generalizing the up-scattering process to new vector and scalar mediators. At this scope, we build upon previous analyses~\cite{Brdar:2018qqj,Li:2020pfy,Chao:2021bvq,Chang:2020jwl,Chen:2021uuw} and consider the possible production of a new MeV-scale dark fermion (DF) at COHERENT. By means of a detailed statistical analysis~\cite{DeRomeri:2022twg}, we combine the most recent CsI and LAr datasets and infer constraints on the DF parameter space. Reference~\cite{Brdar:2018qqj} already considered the possible production of a new fermion at \cevns~experiments, assuming that it couples to neutrinos and quarks via a singlet scalar. In~\cite{Li:2020pfy} loop effects in the up-scattering of neutrinos with \cevns, also mediated by a scalar particle, have been studied. Ref.~\cite{Chang:2020jwl} also studied the possible production of a DF at \cevns~experiments, focusing on dimension-6 effective generalized interactions, analyzing old CsI (2017) data and providing sensitivities for a future LAr detector.  We improve upon these previous results by analyzing new COHERENT datasets (CsI, 2021~\cite{,COHERENT:2021xmm} and LAr, 2020~\cite{COHERENT:2020iec}) and by including timing
information in the data analysis. We also consider the possible production of the DF from the neutrino-electron scattering (ES) on an atomic nucleus, which in some parts of the parameter space drastically enhances the expected signal at the CsI detector.  Going one step beyond Refs.~\cite{Brdar:2018qqj,Chang:2020jwl} we further extend the analysis addressing both (light) scalar and vector mediators. Notice that the possible production of an exotic fermion from neutrino-electron scattering experiments had been studied in~\cite{Chen:2021uuw}, focusing on effective operators. Reference~\cite{Chao:2021bvq} instead considered loop effects on vector and axial-vector contributions in the up-scattering process at COHERENT, using the old CsI data. Finally, the effective operators leading to the up-scattering interactions of interest here have been investigated in the context of DM absorption at direct detection experiments in~\cite{PhysRevLett.124.181301,Dror:2020czw}. In this work, we recompute all the relevant cross sections and provide their full expressions, to be able to explore the explicit dependence of both \cevns~and ES cross sections on the mediator and the DF masses. We conclude with a brief discussion about the stability of the DF.\\

Our paper is organized as follows. In Sec.~\ref{sec:prod} we describe how the dark fermion can be produced through up-scattering, from the interaction of an incoming neutrino with the nuclei or the atomic electrons of the detector. In Sec.~\ref{sec:analysis} we discuss the expected signal and provide details of the statistical analysis. In Sec.~\ref{sec:res} we present our results, in terms of exclusion regions in the dark fermion parameter space. Finally, we summarize our results in Section~\ref{sec:concl}.

\section{Production of a dark fermion at COHERENT}
\label{sec:prod}
We consider the possible production of a DF $\chi$ at COHERENT, from the up-scattering of neutrinos produced at the SNS on the nuclei of the detector:
\begin{equation}
    \nu_\alpha \mathcal{N} \to \chi \mathcal{N} \, , \qquad \nu_\alpha = \nu_\mu,\, \overline{\nu}_\mu,\, \nu_e \, .
\end{equation}

\begin{figure}
    \centering
    \begin{subfigure}{0.4\textwidth}
         \centering
         \includegraphics[width=\textwidth]{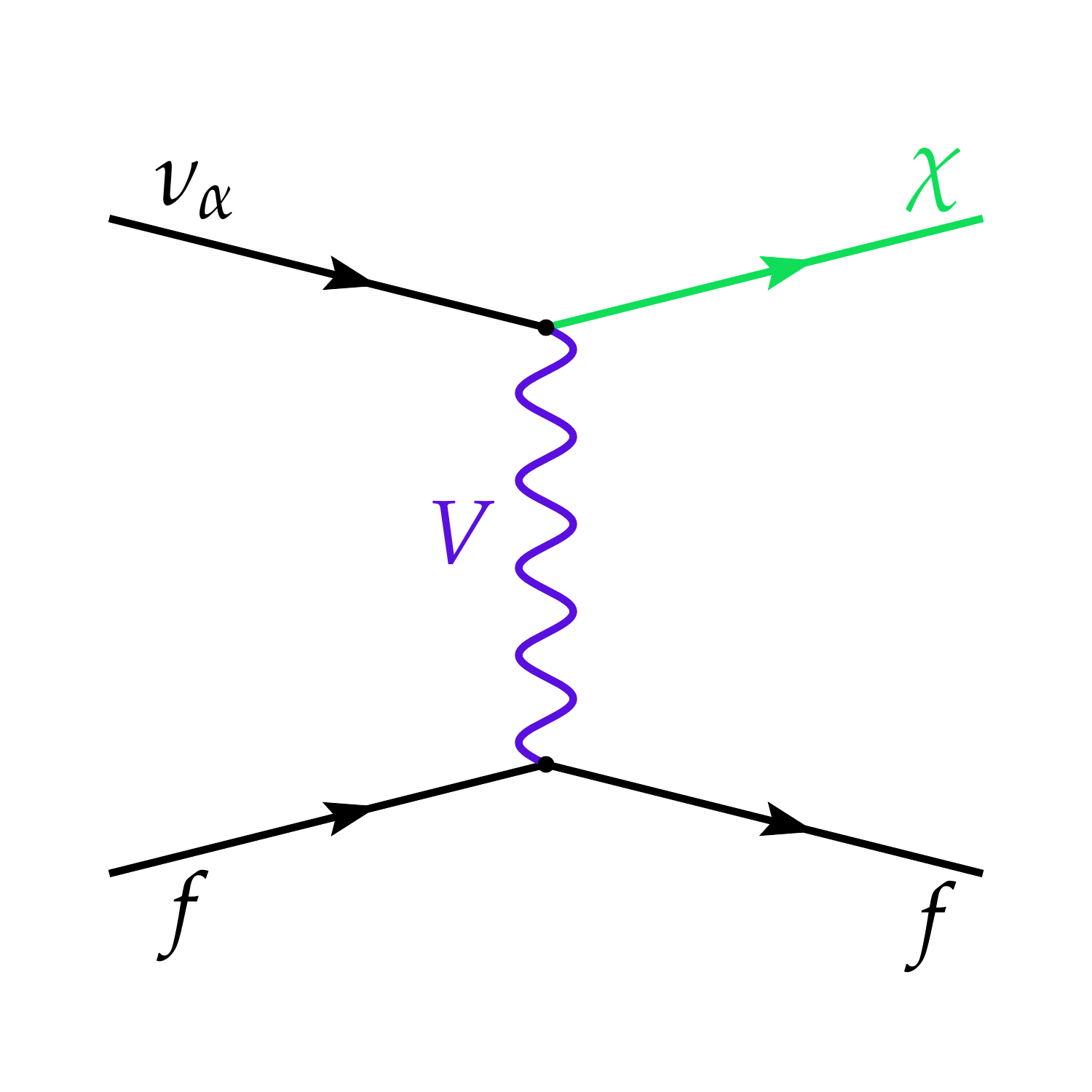}
     \end{subfigure}
     \hspace{-15pt}
     \begin{subfigure}{0.4\textwidth}
         \centering
         \includegraphics[width=\textwidth]{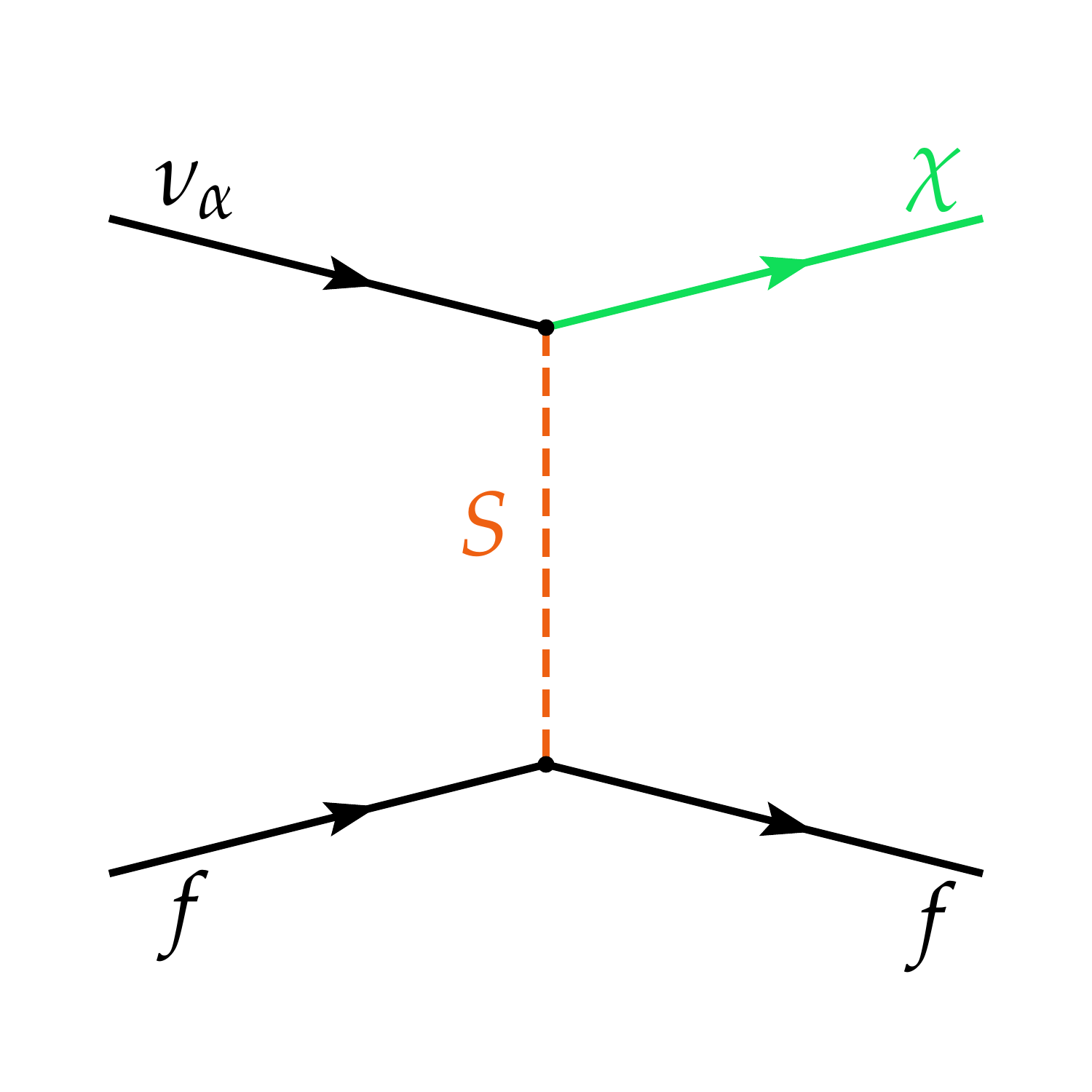}
     \end{subfigure}
    \caption{Feynman diagrams for the $\nu_\alpha f \to \chi f$ process ($f = u,\, d,\, e^-$), mediated by a vector (on the left) or a scalar (on the right) particle. }
    \label{fig:diagrams}
\end{figure}

In the case of the COHERENT-CsI detector, ES events are also important as they can mimic the \cevns~signal, so we also consider the DF production from neutrinos scattering off atomic electrons: 
\begin{equation}
    \nu_\alpha e^- \to \chi e^- \, , \qquad \nu_\alpha = \nu_\mu,\, \overline{\nu}_\mu,\, \nu_e \, .
\end{equation}
The relevant Feynman diagrams are depicted in Fig.~\ref{fig:diagrams}.

In a simplified phenomenological scenario, we assume that at energies below the electroweak scale the SM Lagrangian is extended by 
\begin{eqnarray}
\label{eq:Lagr}
{\mathcal{L}}_\mathrm{DF}^\mathrm{V}&\supseteq& V_\mu\, \overline{\chi} \gamma^\mu \left(g_{\chi_L} P_L + g_{\chi_R} P_R\right) \nu_\alpha + V_\mu\,\sum_f \overline{f} \gamma^\mu \left(g_{f_L} P_L + g_{f_R} P_R\right) f +  \mathrm{H.c.} \, , \nonumber \\
{\mathcal{L}}_\mathrm{DF}^\mathrm{S}&\supseteq& S\, \overline{\chi} \left(g_{\chi_L} P_L + g_{\chi_R} P_R\right) \nu_\alpha + S\,\sum_f \overline{f} \left(g_{f_L} P_L + g_{f_R} P_R\right) f  + \mathrm{H.c.} \, ,
\end{eqnarray}
thus encoding new neutrino-DF interactions mediated by either a real
scalar ($S$) or a vector ($V_\mu$) mediator, also coupling to the charged fermions of the first family, i.e., $f = u,\, d,\, e^-$. 
The strength of the interaction is quantified by the couplings $g_{\chi_L}$ and $g_{\chi_R}$ for the vertex of $\chi$ with the mediator and the neutrino (we assume this coupling to be identical for all neutrino flavors), and $g_{f_L}$ and $g_{f_R}$ for the vertex of $f$ and the mediator. The subscripts $L$ and $R$ in the couplings denote their left- and right-handed components, while $P_L$ and $P_R$ are the left- and right-handed chiral projectors, respectively. Given the nature of neutrinos in the SM, we notice that $g_{\chi_R} = 0$. While in the present work we decide to remain agnostic of the origin of such interactions, we refer the reader to~\cite{Farzan:2018gtr,Chang:2020jwl,
Bertuzzo:2021opb,AtzoriCorona:2022moj} for a discussion on possible UV-completed models leading to such low-scale interactions.
Notice also that we consider the most general case of light mediators, which can give rise to interesting spectral features at \cevns~experiments like COHERENT. As a consequence, a dependence on the mediator mass and the momentum transfer will appear in the up-scattering cross sections, from the propagator. We should mention that several constraints may apply to such light-mediator scenarios, including laboratory limits from fixed-target and beam-dump experiments, rare decays and accelerator data (see for instance~\cite{Bauer:2018onh,Ilten:2018crw}). Most of these constraints require the new mediator to couple to charged leptons and depend on the Lorentz structure of the coupling.  Scalar interactions are also constrained by neutrino masses, to which they can contribute at the loop level. Moreover, light new particles coupling to neutrinos and matter fields can also affect the dynamics of stellar cooling and supernovae~\cite{Farzan:2018gtr,AristizabalSierra:2019ykk} and big bang nucleosynthesis (BBN).\\

Given the Lagrangians in Eq.~(\ref{eq:Lagr}) and the diagrams shown in Fig.~\ref{fig:diagrams}, we can compute the relevant $\nu_\alpha \mathcal{N} \to \chi \mathcal{N}$ cross sections for the vector (V) and scalar (S) mediators, in terms of the nuclear recoil energy $E_\mathrm{nr}$. Note that the following expressions hold for both neutrinos and antineutrinos~\footnote{We have checked that the squared amplitude is the same in both cases, after applying the sum over spins.} and are therefore applicable to all components of the SNS neutrino flux. The up-scattering cross section for the vector case reads
\begin{equation}
    \label{eq:xsec_DF_V}
\begin{aligned}
\frac{\d \sigma_{\nu_\alpha \mathcal{N}}}{\d E_\mathrm{nr}}\Big|^\mathrm{V}_\mathrm{CE \nu NS} (E_\nu, E_{\mathrm{nr}})=& \, \dfrac{9\, m_\mathcal{N}}{4\pi \left(m_{V}^2 + 2 m_\mathcal{N} E_\mathrm{nr}\right)^2} F_W^2(\qtransfer^2) A^2 g_{V}^4 \\[4pt]
& \times \left[\left(2 - \dfrac{m_\mathcal{N} E_\mathrm{nr}}{ E_\nu^2} - \dfrac{2 E_\mathrm{nr}}{E_\nu} + \dfrac{E_\mathrm{nr}^2}{E_\nu^2}\right) - \dfrac{m_\chi^2}{2 E_\nu^2} \left(1 + \dfrac{2 E_\nu}{m_\mathcal{N}} - \dfrac{E_\mathrm{nr}}{m_\mathcal{N}}\right)\right] \, ,
\,
\end{aligned}
\end{equation}
where $E_\nu$ is the incident neutrino energy, $m_\chi$ is the mass of the DF, $m_V$ is the vector mediator mass, $m_\mathcal{N}$ is the nuclear mass and $A$ is the mass number. The vector coupling has been defined as $g_V \equiv \sqrt{g_{\chi_L} g_f}$ (we refer to Appendix~\ref{sec:appendixA} for more details). In the present work, nuclear-physics effects are incorporated through the nuclear form factor $F_W(\qtransfer^2)$, for which we adopt the Klein-Nystrand parametrization, defined as 
\begin{equation}\label{eq:form-factor-kn}
    F_W(\qtransfer^2) = 3\, \dfrac{j_1(\qtransfer R_A)}{\qtransfer R_A (1 + a_k^2 \qtransfer^2)} \, ,
\end{equation}
where $j_1(x) = \sin(x)/x^2 - \cos(x)/x$ is the spherical Bessel function of order one, $\qtransfer \approx \sqrt{2 m_\mathcal{N} E_\mathrm{nr}}$ represents the magnitude of the three-momentum transfer, $R_A = 1.23 \,A^{1/3}~\mathrm{fm}$ is the nuclear radius and $a_k = 0.7~\mathrm{fm}$ is the Yukawa potential range.

The scalar mediator cross section, on the other hand, is given by 
\begin{equation}
    \label{eq:xsec_DF_S}
\begin{aligned}
\frac{\d \sigma_{\nu_\alpha \mathcal{N}}}{\d E_\mathrm{nr}}\Big|_\mathrm{CE\nu NS}^\mathrm{S} (E_\nu, E_{\mathrm{nr}}) =& \, \dfrac{m_\mathcal{N}}{8\pi \left(m_{S}^2 + 2 m_\mathcal{N} E_\mathrm{nr}\right)^2} C_{S}^4 F_W^2 (\qtransfer^2) \\[4pt]
& \times \left(2 + \dfrac{E_\mathrm{nr}}{m_\mathcal{N}}\right)\left(\dfrac{m_\mathcal{N}E_\mathrm{nr}}{E_\nu^2} + \dfrac{m_\chi^2}{2 E_\nu^2}\right)
  \, .
\,
\end{aligned}
\end{equation}
where $m_S$ is is the scalar mediator mass and $C_S$ is the scalar coupling. It is defined as
\begin{equation}\label{eq:scalar_coupling}
C_S^2 \equiv g_S^2 \left( Z \sum_{q = u, d} \dfrac{m_p}{m_q} f_{T_q}^{(p)} + N \sum_{q = u,d} \dfrac{m_n}{m_q} f_{T_q}^{(n)} \right),
\end{equation}
where $g_S \equiv \sqrt{g_{\chi_L} g_f}$, $Z$ is the number of protons of the nucleus and $N = A - Z$ is the number of neutrons, $m_p$ and $m_n$ are the proton and neutron masses, respectively, and $m_q$ is the quark mass. Here, $f_{T_q}^{(p)}$ and $f_{T_q}^{(n)}$ indicate the quark-mass contributions to the nucleon (proton and neutron) mass. While observing that the determination of these values is still uncertain~\cite{Cirelli:2013ufw, DelNobile:2021wmp},
in the following we fix~\cite{DelNobile:2021wmp}
\begin{align}\label{eq:quark-mass-f-values}
    f_{T_u}^{(p)} &= 0.026 \,, \qquad f_{T_u}^{(n)} = 0.018 \,, \\
    f_{T_d}^{(p)} &= 0.038 \,, \qquad f_{T_d}^{(n)} = 0.056 \,.
\end{align}
We have checked that using a different value~\cite{Ellis:2018dmb}  would change the value of $C_S^2$ by a factor of $\sim 30 \%$, at most, however the sensitivity on $g_S$ remains essentially unaffected.

Possible axial-vector and pseudoscalar contributions are in general suppressed by the nuclear spin as detailed in Ref.~\cite{AristizabalSierra:2018eqm}. Let us also note that in the present analysis, the left- and right-handed DF couplings are taken to be equal in Eq.~\eqref{eq:Lagr}, thus axial-vector and pseudoscalar contributions are vanishing (see Appendix~\ref{sec:appendixA}).\\

Next, we provide expressions for the production of a DF via ES on an atomic nucleus $\mathcal{A}$ containing $Z$ protons. Since the electrons are bound in $\mathcal{A}$, for a given energy deposition $E_\mathrm{er}$ the relevant cross sections are expressed as the product of the free ES cross section times the effective charge $Z_{\text{eff}}^{\mathcal{A}}(E_\mathrm{er})$, as
\begin{equation}
    \label{eq:xsec_DF_ES_V}
\begin{aligned}
\frac{\d \sigma_{\nu_\alpha \mathcal{A}}}{\d E_\mathrm{er}}\Big|_\mathrm{ES}^\mathrm{V} (E_\nu, E_{\mathrm{er}})=& \, Z_{\mathrm{eff}}^{\mathcal{A}}\left(E_{\mathrm{er}}\right) \dfrac{m_e g_V^4}{4\pi \left(m_V^2 + 2 m_e E_{\mathrm{er}}\right)^2}  \\[4pt]
& \times \left[\left(2 - \dfrac{m_e E_{\mathrm{er}}}{ E_\nu^2} - \dfrac{2 E_{\mathrm{er}}}{E_\nu} + \dfrac{E_{\mathrm{er}}^2}{E_\nu^2}\right) - \dfrac{m_\chi^2}{2 E_\nu^2} \left(1 + \dfrac{2 E_\nu}{m_e} - \dfrac{E_{\mathrm{er}}}{m_e}\right)\right] \, ,
\,
\end{aligned}
\end{equation}

\begin{equation}
    \label{eq:xsec_DF_ES_S}
\frac{\d \sigma_{\nu_\alpha \mathcal{A}}}{\d E_\mathrm{er}}\Big|_\mathrm{ES}^\mathrm{S} (E_\nu, E_{\mathrm{er}}) = \, Z_{\mathrm{eff}}^{\mathcal{A}}\left(E_{\mathrm{er}}\right) \dfrac{m_e g_S^4}{8\pi \left(m_S^2 + 2 m_e E_{\mathrm{er}}\right)^2}   \left(2 + \dfrac{E_{\mathrm{er}}}{m_e}\right) \left(\dfrac{m_\chi^2}{2 E_\nu^2} + \dfrac{m_e E_{\mathrm{er}}}{E_\nu^2}\right)
  \, ,
\,
\end{equation}
where $m_e$ is the electron mass. The effective charges $Z_{\text{eff}}^{\mathcal{A}}(E_\mathrm{er})$ for Cs and I isotopes are given in Appendix~\ref{sec:appendixB}.

A few comments are in order. First, let us note that in the limit  $m_\chi \to 0$ one recovers exactly the expressions for \cevns ~and ES with new light mediators (see for instance~\cite{DeRomeri:2022twg,Cerdeno:2016sfi,Majumdar:2021vdw}). Second, throughout the manuscript we will assume that the new mediators couple equally to either $u,\, d$ or  $e^-$. The hypothesis of universal couplings reduces the degrees of freedom  simplifying the analysis, thus allowing us to express the results in terms of a single coupling, namely $g_\mathrm{S}$ or $g_\mathrm{V}$, and the corresponding mediator and DF masses. While this model is anomalous, it can be made anomaly-free by introducing new particles, which may also belong to the dark sector~\cite{AtzoriCorona:2022moj}. Moreover, while the scalar and vector DF cross sections can in principle involve flavor-dependent terms, in the present work we restrict to DF-production cross sections which are flavor-blind. It is finally important to note that the DF mass is constrained from above: $m_\chi \leq \sqrt{m\left(m_\mu + m\right)} - m$, where $m$ is the mass of the fixed target ($m_\mathcal{N}$ or $m_e$) and $m_\mu$ is the muon mass. This bound is dictated by kinematics and implies that the process shown in Fig.~\ref{fig:diagrams} will not occur if $m_\chi$ is above that limit. 
For the sake of completeness, let us conclude mentioning that the SM \cevns ~and ES cross sections are well-known results in the literature and can be found in, e.g., Refs.~\cite{Abdullah:2022zue} and~\cite{Giunti:2007ry}. In the SM expressions for both \cevns~and ES, the $V-A$ interference term appears with a different sign for neutrino or antineutrino scattering. However, in the case of \cevns, the axial contributions are tiny for the nuclei of interest in the present work and hence neglected in our calculations~\footnote{In the SM case, angular momentum conservation implies that the axial-vector contribution to the  \cevns ~cross section vanishes for even-even nuclei, while for the other cases the contribution is tiny, especially for heavy nuclei~\cite{Barranco:2005yy}.}. Moreover, while in the SM the \cevns ~cross section is flavor-blind, the SM ES cross section is different for $\nu_e-e^-$ compared to $\nu_{\mu, \tau}-e^-$ scattering. The reason is that the former acquires contributions from both charged- and neutral-currents, while for the latter only neutral-currents are relevant. Before closing our discussion let us clarify that there is no interference between the vector/scalar \cevns ~cross sections and the SM one. The same holds for the ES contribution.

\section{Data analysis}
\label{sec:analysis}
We now proceed to analyze the most recent COHERENT datasets, from the CsI~\cite{COHERENT:2021xmm} and LAr~\cite{COHERENT:2020ybo} detectors. In each case, we take into account energy and timing information as well as all relevant systematic effects, as done in Ref.~\cite{DeRomeri:2022twg}. Moreover, in the CsI case, we also include possible contributions from ES events~\cite{Coloma:2022avw, AtzoriCorona:2022qrf,DeRomeri:2022twg}, which turn out to be relevant especially in the vector-mediated scenario. In the analysis of the LAr  dataset, the measurement of the ratio of the integrated photomultiplier amplitude and the total amplitude
 in the first 90~ns ($F_{90}$) allows to distinguish between \cevns-induced nuclear recoils and ES-induced electron recoils and therefore we do not include the ES contribution.
 
For the COHERENT-CsI (COHERENT-LAr) analysis, we assume a detector mass $m_{\rm det} = 14.6 ~(24)$ kg located at a distance $L=19.3 ~(27.5)$~m from the SNS source. For the calculation of \cevns~events, $N_{ij}^{\mathrm{CE}\nu\mathrm{NS}}(\mathcal{N})$, we follow the procedure described in Ref.~\cite{DeRomeri:2022twg}. The expected number of DF or SM events, on a nuclear target $\mathcal{N}$, per neutrino flavor,  $\nu_\alpha$, and in each nuclear recoil energy bin $i$ is given by
\begin{align}
\label{eq:Nevents_DF_CEvNS}
N_{i, \nu_\alpha, \mathrm{\kappa}}^{\mathrm{CE\nu NS}}(\mathcal{N})
= \nonumber
& \, N_\mathrm{target}
\int_{E_{\mathrm{nr}}^{i}}^{E_{\mathrm{nr}}^{i+1}}
\hspace{-0.3cm}
\d E_{\mathrm{nr}}\,
\epsilon_E(E_{\mathrm{nr}})
\int_{E^{\prime\mathrm{min}}_{\mathrm{nr}}}^{E^{\prime\text{max}}_{\text{nr}}}
\hspace{-0.3cm}
\d E'_{\text{nr}}
\,
\mathcal{R}(E_{\text{nr}},E'_{\text{nr}})  \\
& \times \int_{E_\nu^{\text{min}}(E'_{\text{nr}})}^{E_\nu^{\text{max}}}
\hspace{-0.3cm}
\d E_\nu \,
\frac{\d N_{\nu_\alpha}}{\d E_\nu}(E_\nu)
\frac{\d \sigma_{\nu_\alpha \mathcal{N}}}{\d E'_\mathrm{nr}}\Big|_\mathrm{CE\nu NS}^\mathrm{\kappa}(E_\nu, E'_{\mathrm{nr}})
,
\end{align}
where $N_\mathrm{target} = N_{\mathrm{A}} m_{\mathrm{det}} / M_{\mathrm{\mathrm{target}}}$ is the number of target atoms in the detector, with $N_{\mathrm{A}}$ being the Avogadro's constant and $M_{\mathrm{\mathrm{target}}}$ the molar mass of the detector material. Index $\mathrm{\kappa}$ accounts for the different interactions, namely $\mathrm{\kappa= \{SM, S, V\}}$. The three components of the differential $\pi$-DAR neutrino flux produced at the SNS read
\begin{equation}
\begin{aligned} 
\frac{\d N_{\nu_\mu}}{\d E_\nu}(E_\nu) & = \eta \, \delta\left(E_\nu-\frac{m_{\pi}^{2}-m_{\mu}^{2}}{2 m_{\pi}}\right) \quad &(\text{prompt})\, , \\ 
\frac{\d N_{\bar{\nu}_\mu}}{\d E_\nu}(E_\nu) & = \eta \frac{64 E^{2}_\nu}{m_{\mu}^{3}}\left(\frac{3}{4}-\frac{E_\nu}{m_{\mu}}\right) \quad &(\text{delayed})\, ,\\ 
\frac{\d N_{\nu_e}}{\d E_\nu}(E_\nu) & = \eta \frac{192 E^{2}_\nu}{m_{\mu}^{3}}\left(\frac{1}{2}-\frac{E_\nu}{m_{\mu}}\right) \quad &(\text{delayed}) \, ,
\end{aligned}
\label{labor-nu}
\end{equation}
($m_{\pi}$ is the pion mass) and are normalized to $\eta = r N_{\mathrm{POT}}/4 \pi L^2$, where $r$ denotes the number of neutrinos per flavor produced for each proton on target (POT). For the CsI detector $r=0.0848$ and $N_{\mathrm{POT}}=3.198  \times 10^{23}$, while for the LAr detector $r=0.009$ and $N_{\mathrm{POT}}=1.38 \times 10^{23}$. The integration limits read
\begin{align}
\label{eq:energy-limits-dark-fermion}
    &E_{\nu}^{\mathrm{min}} (E^{\prime}_{\mathrm{nr}}) = \dfrac{2m_{\mathcal{N}} \left(E^{\prime}_{\mathrm{nr}} \right)^2 + m_\chi^2 E^{\prime}_{\mathrm{nr}}}{4m_{\mathcal{N}}E^{\prime}_{\mathrm{nr}}} \nonumber \\[4pt]
    &\hspace{5em}+ \dfrac{\sqrt{\left(2m_{\mathcal{N}} \left(E^{\prime}_{\mathrm{nr}} \right)^2 + m_\chi^2 E^{\prime}_{\mathrm{nr}}\right)^2 + 2m_{\mathcal{N}}E^{\prime}_{\mathrm{nr}}\left(m_\chi^4 + 4m_{\mathcal{N}}^2 \left(E^{\prime}_{\mathrm{nr}} \right)^2 + 4m_{\mathcal{N}} m_\chi^2 E^{\prime}_{\mathrm{nr}}\right)}}{4m_{\mathcal{N}}E^{\prime}_{\mathrm{nr}}} \nonumber 
    \\[4pt]
    &\hspace{5em} \approx \sqrt{\dfrac{m_{\mathcal{N}}E^{\prime}_{\mathrm{nr}}}{2}}\left(1 + \dfrac{m_\chi^2}{2m_{\mathcal{N}}E^{\prime}_{\mathrm{nr}}}\right), \nonumber \\[4pt]
    &E_{\nu}^{\mathrm{max}} = \dfrac{m_\mu}{2}, \nonumber \\[4pt]
    &E^{\prime \substack{\scriptscriptstyle \mathrm{min} \\ \scriptscriptstyle \mathrm{(max)}}}_{\mathrm{nr}}  = \dfrac{2m_{\mathcal{N}}{(E_{\nu}^{\mathrm{max}})}^{2} - m_\chi^2\left(E_{\nu}^{\mathrm{max}} + m_{\mathcal{N}}\right) {\substack{- \\ (+)}} E_{\nu}^{\mathrm{max}}\sqrt{4m_{\mathcal{N}}^2 {(E_{\nu}^\mathrm{max})}^{2} - 4m_{\mathcal{N}} m_\chi^2 \left(E_{\nu}^{\mathrm{max}} + m_{\mathcal{N}}\right) + m_\chi^4}}{2m_{\mathcal{N}}\left(2E_{\nu}^{\mathrm{max}} + m_{\mathcal{N}}\right)}.
\end{align}
Notice that when $m_\chi \to 0$ the integration limits relevant to the SM case are recovered. The remaining detector-specific quantities, namely the energy resolution function $\mathcal{R}(E_{\text{nr}},E'_{\text{nr}})$  relating the true nuclear recoil energy ($E'_{\text{nr}}$) with the reconstructed one ($E_{\text{nr}}$) as well as the  energy-dependent detector efficiency $\epsilon_E(E_{\mathrm{nr}})$ are explained in detail in Appendix~\ref{sec:appendixC}.

We further include timing information in the analysis, by distributing the predicted $N_{i, \nu_{\alpha}, \mathrm{\kappa}}^{\mathrm{CE\nu NS}}(\mathcal{N})$ in each time bin $j$. We use the time distributions $\mathcal{P}^{\nu_\alpha}_T(t_{\mathrm{rec}})$ of $\nu_\alpha = \nu_\mu,\, \overline{\nu}_\mu,\, \nu_e$ given in~\cite{Picciau:2022xzi,COHERENT:2021xmm}, normalized to 6 $\mathrm{\mu s}$. Finally, the predicted  event number, per observed nuclear recoil energy and time bins $i, j$ is given by
\begin{equation}
\label{eq:Nevents_DF_ij_CEvNS}
 N_{ij,\mathrm{\kappa}}^{\mathrm{CE\nu NS}}(\mathcal{N}) = \sum_{\nu_\alpha =\nu_{e}, \nu_{\mu}, \bar{\nu}_{\mu}}\int_{t_{\mathrm{rec}}^{j}}^{t_{\mathrm{rec}}^{j+1}} \d t_{\mathrm{rec}} \, \mathcal{P}^{\nu_\alpha}_T(t_{\mathrm{rec}}, \alpha_6)\epsilon_T(t_{\mathrm{rec}}) N_{i, \nu_{\alpha},\mathrm{\kappa}}^{\mathrm{CE\nu NS}}(\mathcal{N}),
\end{equation}
where $\epsilon_T (t_{\mathrm{rec}})$ is the time-dependent efficiency. Notice that we include an additional nuisance parameter on the beam timing, i.e. $\alpha_6$~\cite{DeRomeri:2022twg} (for details, see the discussion below and in Appendix~\ref{sec:appendixC}). Then, the total \cevns ~event rate is simply given by 
\begin{equation}
\label{eq:total_CEvNS}
\begin{aligned}
      N_{ij}^{\mathrm{CE\nu NS}} = & \sum_{\mathrm{\kappa=SM, V}} \sum_{\mathcal{N}} N_{ij,\mathrm{\kappa}}^{\mathrm{CE\nu NS}}(\mathcal{N}) \qquad & \text{(vector mediator),}\\
      N_{ij}^{\mathrm{CE\nu NS}} = & \sum_{\mathrm{\kappa=SM, S}} \sum_{\mathcal{N}} N_{ij,\mathrm{\kappa}}^{\mathrm{CE\nu NS}}(\mathcal{N}) \qquad & \text{(scalar mediator),}
      \end{aligned}
\end{equation}
where the inner sum runs over the Cs and I isotopes for the case of CsI detector, while for the case of LAr the inner sum is trivially dropped.

Similarly to  the \cevns ~case described above,  the expected number of DF (or SM) ES events, on an atomic nucleus $\mathcal{A}$, per neutrino flavor,  $\nu_\alpha$, and in each electron recoil energy bin $i$  is given by
\begin{align}
\label{eq:Nevents_DF_ES}
N_{i, \nu_\alpha, \mathrm{\kappa}}^{\mathrm{ES}}(\mathcal{A})
= \nonumber
& \, N_\mathrm{target}
\int_{E_{\mathrm{er}}^{i}}^{E_{\mathrm{er}}^{i+1}}
\hspace{-0.3cm}
\d E_{\mathrm{er}}\,
\epsilon_E(E_{\mathrm{er}})
\int_{E^{\prime\mathrm{min}}_{\mathrm{nr}}}^{E^{\prime\text{max}}_{\text{er}}}
\hspace{-0.3cm}
\d E'_{\text{er}}
\,
\mathcal{R}(E_{\text{er}},E'_{\text{er}})  \\
& \times \int_{E_\nu^{\text{min}}(E'_{\text{er}})}^{E_\nu^{\text{max}}}
\hspace{-0.3cm}
\d E_\nu \, 
\frac{\d N_{\nu_\alpha}}{\d E_\nu}(E_\nu)
\frac{\d \sigma_{\nu_\alpha \mathcal{A}}}{\d E'_\mathrm{er}}\Big|_\mathrm{ES}^\mathrm{\kappa}(E_\nu, E'_{\mathrm{er}}) \, .
\end{align}
The nuclear recoil energy and the corresponding electron-equivalent energy are related via the quenching factor (QF) (see Appendix~\ref{sec:appendixC} for details), while the integration limits are given by Eq.~(\ref{eq:energy-limits-dark-fermion}) via the substitutions $m_\mathcal{N} \to m_e$ and $E'_\text{nr} \to E'_\text{er}$~\footnote{The approximated expression given for $E_\nu^\text{min}(E'_\text{nr})$ in Eq.~\eqref{eq:energy-limits-dark-fermion} is only valid for \cevns, not for ES.}. Then, the corresponding  event rate per observed electron recoil energy $i$ and time bin $j$ is given by 
\begin{equation}
\label{eq:Nevents_DF_ij_ES}
 N_{ij,\mathrm{\kappa}}^{\mathrm{ES}}(\mathcal{A}) = \sum_{\nu_\alpha =\nu_{e}, \nu_{\mu}, \bar{\nu}_{\mu}}\int_{t_{\mathrm{rec}}^{j}}^{t_{\mathrm{rec}}^{j+1}} \d t_{\mathrm{rec}} \, \mathcal{P}^{\nu_\alpha}_T(t_{\mathrm{rec}}, \alpha_6)\epsilon_T(t_{\mathrm{rec}}) N_{i, \nu_{\alpha},\mathrm{\kappa}}^{\mathrm{ES}}(\mathcal{A}) \, .
\end{equation}
Finally, the total ES event rate is simply given by 
\begin{equation}
\begin{aligned}
      N_{ij}^{\mathrm{ES}} = & \sum_{\mathrm{\kappa=SM, V}} \sum_{\mathcal{N}} N_{ij,\mathrm{\kappa}}^{\mathrm{ES}}(\mathcal{A}) \qquad &\text{(vector mediator),}\\
      N_{ij}^{\mathrm{ES}} = & \sum_{\mathrm{\kappa=SM, S}} \sum_{\mathcal{N}} N_{ij,\mathrm{\kappa}}^{\mathrm{ES}}(\mathcal{A}) \qquad & \text{(scalar mediator),}
      \end{aligned}
\end{equation}
see the discussion below Eq.~(\ref{eq:total_CEvNS}).

For the statistical analysis of COHERENT data we rely on the following Poissonian least-squares function~\cite{DeRomeri:2022twg} 
\begin{equation}
	\chi^2_{\mathrm{CsI}}\Big|_{\mathrm{CE\nu NS} + \mathrm{ES}}
	 =
	2
	\sum_{i=1}^{9}
	\sum_{j=1}^{11}
	\left[ N^\mathrm{th}_{ij}  -  N_{ij}^{\text{exp}} 
	 +  N_{ij}^{\text{exp}} \ln\left(\frac{N_{ij}^{\text{exp}}}{ N^\mathrm{th}_{ij}} \right)\right]\\
	+ \sum_{k=0}^{5}
	\left(
	\dfrac{ \alpha_{k} }{ \sigma_{k} }
	\right)^2  
	.
	\label{chi2CsI}
\end{equation}
The predicted number of events, including \cevns, DF and background events, is defined as 
\begin{align}\nonumber
N^\mathrm{th}_{ij} = &(1 + \alpha_{0} +\alpha_{5})  N_{ij}^{\mathrm{CE\nu NS}} (\alpha_{4}, \alpha_{6}, \alpha_{7})  + (1 + \alpha_{0} )  N_{ij}^{\mathrm{ES}} (\alpha_{6}, \alpha_{7})\\[4pt]
 &+ (1 + \alpha_{1}) N_{ij}^\mathrm{BRN}(\alpha_{6}) + (1 + \alpha_{2}) N_{ij}^\mathrm{NIN}(\alpha_{6})
 + (1 + \alpha_{3}) N_{ij}^\mathrm{SSB}  \,.
	\label{eq:Nth_CsI_chi2}
\end{align}

These expressions involve several nuisances ($\alpha_i$) together with their associated uncertainties ($\sigma_i$). In particular, $\sigma_{0} = 11\%$  encodes efficiency and flux uncertainties; $\sigma_{1} = 25\%$, $\sigma_{2} = 35\%$ and $\sigma_{3} = 2.1\%$ are related to the backgrounds, beam related neutrons (BRN), neutrino induced neutrons (NIN) and  steady state background (SSB), respectively. Finally,
$\sigma_{5} = 3.8\%$ is associated to the QF. Notice that the number of events $N_{ij}^{\mathrm{CE\nu NS}}$, $N_{ij}^{\mathrm{DF}}$ and $N_{ij}^{\mathrm{ES}}$ also include nuisance parameters: $\alpha_{4}$, which enters the nuclear form factor and thus affects only the \cevns ~number of events~\footnote{This is done by introducing the nuisance parameter $\alpha_4$ in the nuclear radius entering Eq.~(\ref{eq:form-factor-kn}) such that $R_A = 1.23 \, A^{1/3} (1+\alpha_4)$.}, with  $\sigma_{4} = 5\%$; $\alpha_{6}$ accommodates the uncertainty in beam timing with no prior assigned, while $\alpha_{7}$ allows for deviations of the uncertainty in the \cevns ~efficiency. We refer the reader to Ref.~\cite{DeRomeri:2022twg} for further details about the statistical analysis.

For the COHERENT-LAr $\chi^2$ analysis we instead consider the following Gaussian least-squares function, based on~\cite{AtzoriCorona:2022moj,DeRomeri:2022twg} 
\begin{equation}
	\chi^2_{\mathrm{LAr}} = 
	\sum_{i=1}^{12}
	\sum_{j=1}^{10}
	\left( \frac{N^\mathrm{th}_{ij}  -  N_{ij}^{\text{exp} } }{\sigma_{ij}}  \right)^2  + \sum_{k=0,3,4,8}
	\left( 	\dfrac{ \beta_{k} }{ \sigma_{k} } 	\right)^2  +  \sum_{k=1,2,5,6,7} \left(\beta_{k}  	\right)^2 \,  ,
\label{chi2LAr}
\end{equation}
with the experimental uncertainty being $\sigma_{ij}^2 = N_{ij}^\text{exp} +  N_{ij}^\mathrm{SSB}/5$, while the theoretical signal is calculated as
\begin{equation}
    \begin{aligned}
     N^\mathrm{th}_{ij} =& (1 + \beta_0 +  \beta_1 \Delta_{\mathrm{CE\nu NS}}^{F_{90+}} + \beta_1 \Delta_{\mathrm{CE\nu NS}}^{F_{90-}} + \beta_2 \Delta_{\mathrm{CE\nu NS}}^\mathrm{t_{trig}}) N_{ij}^{\mathrm{CE\nu NS}}   + (1 + \beta_3) N_{ij}^\mathrm{SSB}  \\
    & + (1 + \beta_4 + \beta_5 \Delta_\mathrm{pBRN}^{E_+} + \beta_5 \Delta_\mathrm{pBRN}^{E_-}
    + \beta_6 \Delta_\mathrm{pBRN}^{t_\text{trig}^+} + \beta_6 \Delta_\mathrm{pBRN}^{t_\text{trig}^-} + \beta_7 \Delta_\mathrm{pBRN}^{t_\text{trig}^\text{w}}) N_{ij}^\mathrm{pBRN}\\
    & + (1 + \beta_8) N_{ij}^\mathrm{dBRN} \, .
    \end{aligned}
\end{equation}
Here, the nuisance parameters $\beta_0,~\beta_3,~\beta_4$ and $\beta_8$ account for the normalization uncertainties of \cevns, SS, prompt BRN (pBRN) and delayed BRN (dBRN) rates, with $\{\sigma_{0},~\sigma_{3},~\sigma_{4},~\sigma_{8}\}$ = $\{0.13,~0.0079,~0.32,~1.0\}$~\cite{COHERENT:2020iec}. Relevant systematic effects affecting the shape uncertainties of the \cevns ~ and pBRN rates are also taken into account through the nuisance parameters $\beta_1, \beta_2, \beta_5, \beta_6$ and $\beta_7$. In particular, $\beta_1$ and $\beta_2$ account for the uncertainty on the \cevns ~shape due to existing systematic uncertainties on the $\pm 1\sigma$ energy distributions of the $F_{90}$ parameter ($\Delta_{\mathrm{CE\nu NS}}^{F_{90\pm}}$) and due to the mean time to trigger distribution ($\Delta_{\mathrm{CE\nu NS}}^\mathrm{t_{trig}}$), respectively.  On the other hand $\beta_5$, $\beta_6$ and $\beta_7$ are introduced to quantify the pBRN shape uncertainty due to the corresponding uncertainty on the  $\pm 1\sigma$ energy distributions ($\Delta_\mathrm{pBRN}^{E_\pm}$),  the $\pm 1 \sigma$ mean time to trigger distributions ($\Delta_\mathrm{pBRN}^{t_\text{trig}^\pm}$) as well as the 
trigger width distribution ($\Delta_\mathrm{pBRN}^{t_\text{trig}^\text{w}}$). The latter five distributions are defined as
\begin{equation}
    \Delta_{\lambda}^{\xi_\lambda} = \frac{N_{ij}^{\lambda,\xi_\lambda} - N_{ij}^{\lambda,\mathrm{CV}}}{N_{ij}^{\lambda,\mathrm{CV}}} \, ,
    \label{eq:Deltas}
\end{equation}
with $\lambda=$ \{CE$\nu$NS, pBRN\} and $\xi_\lambda$ corresponding to the different source uncertainties affecting the \cevns ~or  pBRN shapes. Ιn the definition of Eq.~(\ref{eq:Deltas}) the superscript ``CV"  denotes the central values of the \cevns ~or
pBRN distributions, available in the COHERENT-LAr data release~\cite{COHERENT:2020ybo}. Before closing this discussion, let us note that the $\beta_0$ component includes several uncertainties accounting for the flux (10\%), efficiency (3.6\%), energy calibration (0.8\%), the calibration of the pulse-shape discrimination parameter $F_{90}$  (7.8\%), QF (1\%), and nuclear form factor (2\%)~\cite{COHERENT:2020iec}.

\section{Results}
\label{sec:res}
In this section, we present the obtained constraints on the DF parameter space from the combined analysis of COHERENT data from the CsI~\cite{COHERENT:2021xmm} and LAr~\cite{COHERENT:2020iec} detectors. We start discussing the results for the vector-mediator case and then we proceed to the scalar one. We further compare our results with existing limits in the literature (when applicable). We finally briefly comment about the stability of the DF and a possible connection to the dark matter.

\subsection{Vector mediator}
\label{subsec:results-vector-mediator}
In the following, we refer to the Lagrangian given in Eq.~\eqref{eq:Lagr} and to the cross sections given in Eqs.~\eqref{eq:xsec_DF_V} and \eqref{eq:xsec_DF_ES_V}. Thanks to the redefinition of the coupling $g_V \equiv \sqrt{g_{\chi_L} g_f}$, with $f = u,\,d$ for \cevns~and $f=e^-$ for ES, we can express our results in terms of the relevant parameters, namely the DF mass ($m_\chi$), the mediator mass ($m_V$) and the coupling ($g_V$). This simplification is based on the assumption that the vector mediator has universal couplings to both quarks and electrons.

\begin{figure}
\centering
\begin{subfigure}{0.49\textwidth}
    \includegraphics[width=\textwidth]{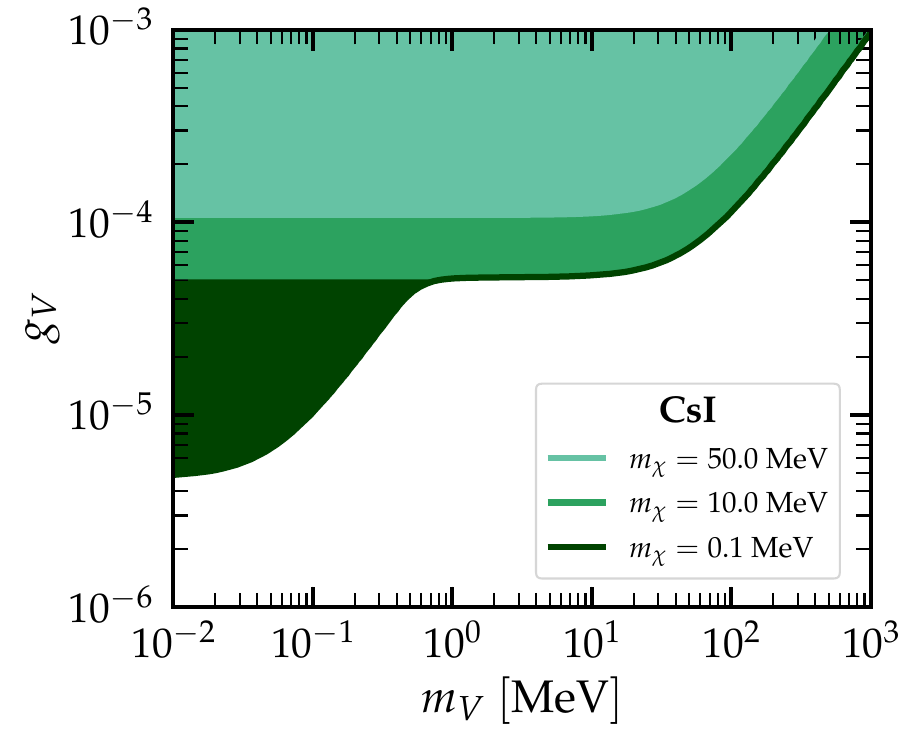}
\end{subfigure}
\hfill
\begin{subfigure}{0.49\textwidth}
    \includegraphics[width=\textwidth]{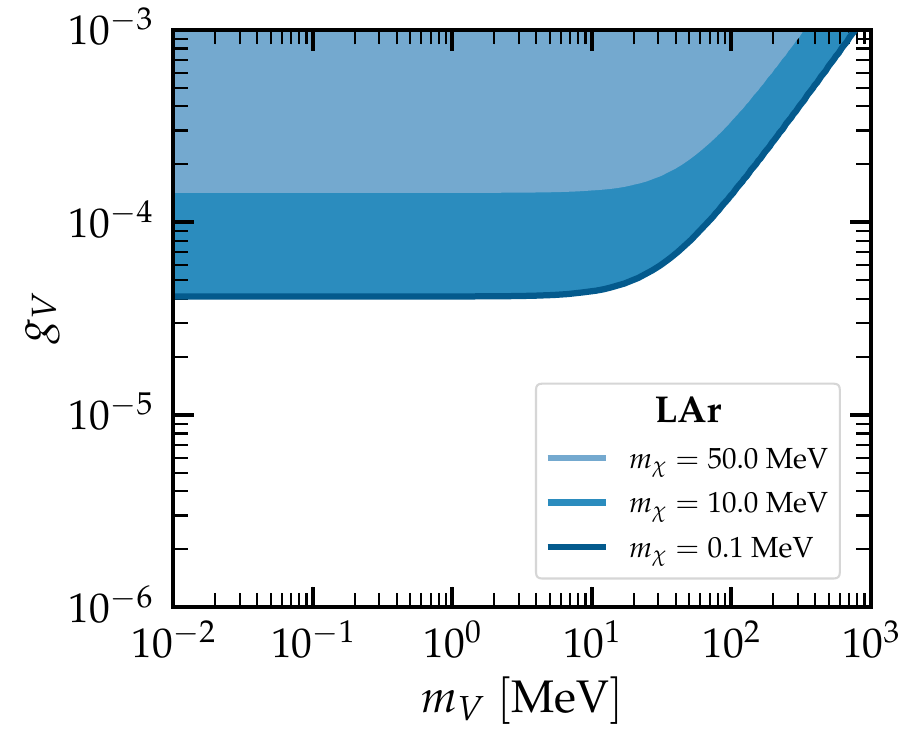}
\end{subfigure}

\begin{subfigure}{0.49\textwidth}
    \includegraphics[width=\textwidth]{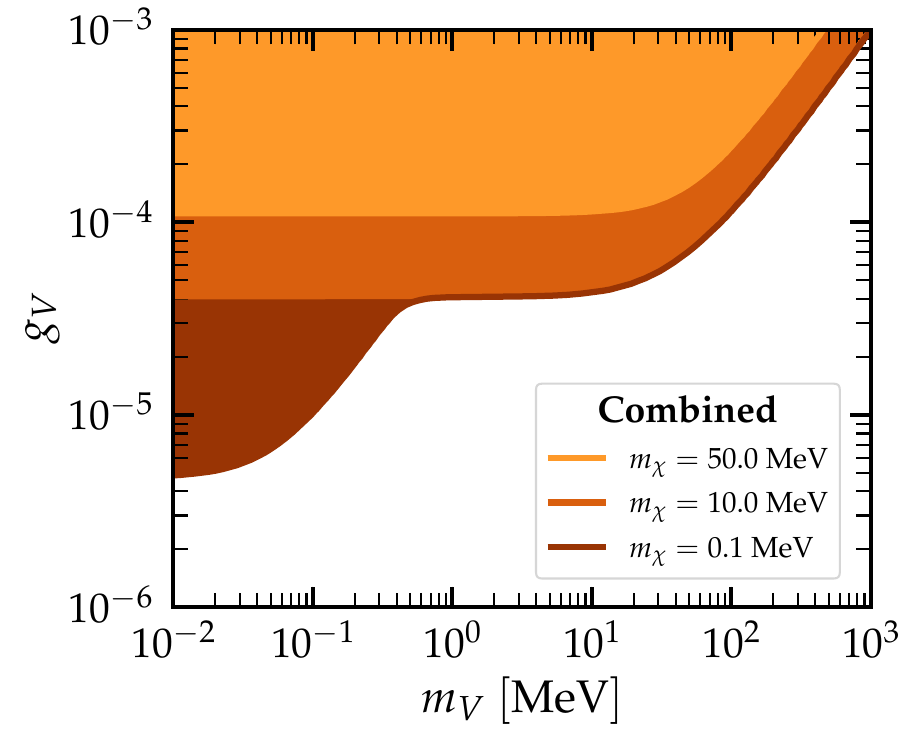}
\end{subfigure}
\hfill
\begin{subfigure}{0.49\textwidth}
    \includegraphics[width=\textwidth]{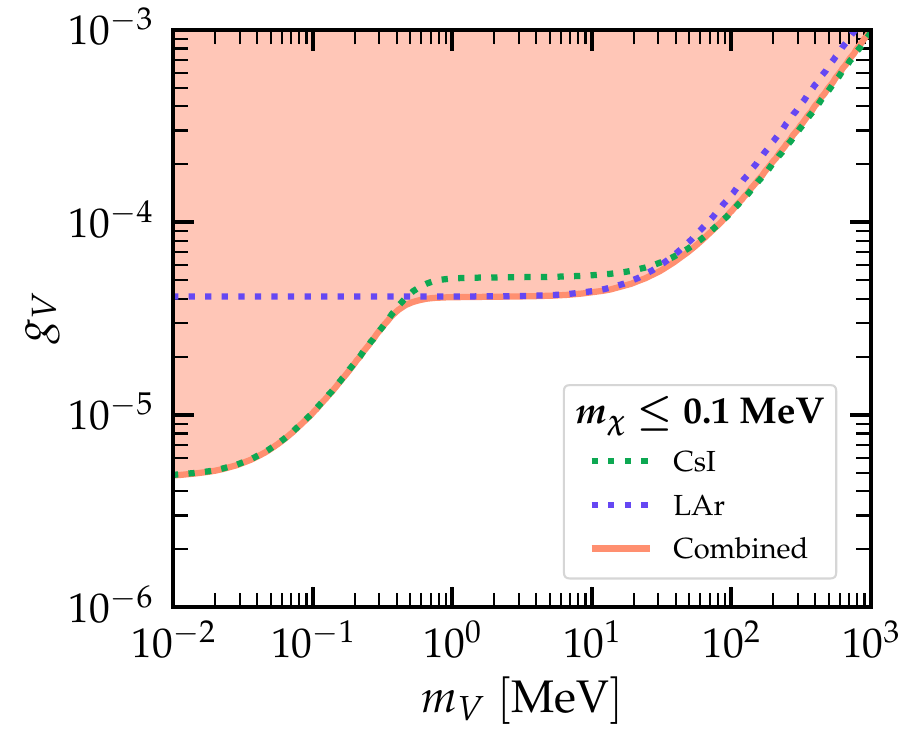}
\end{subfigure}
\caption{$90\%$ C.L. (2 d.o.f.) exclusion regions in the $m_V-g_V$ plane, for the vector-mediator case. Different fixed values of $m_\chi$ are considered. The top row shows the analysis done with the CsI (left) and LAr (right) COHERENT  data, while the bottom row shows the combined analysis (left) and a comparison between the different analyses for the most constraining case (right).}
\label{fig:fixed-mchi-vector-mediator}
\end{figure}

\begin{figure}
\centering
\begin{subfigure}{0.49\textwidth}
    \includegraphics[width=\textwidth]{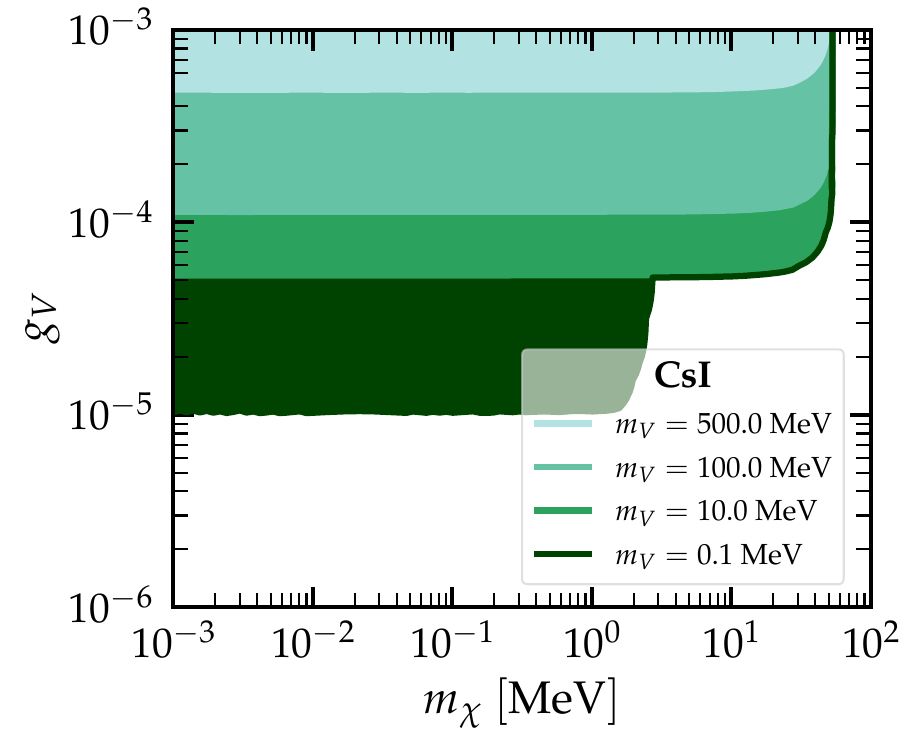}
\end{subfigure}
\hfill
\begin{subfigure}{0.49\textwidth}
    \includegraphics[width=\textwidth]{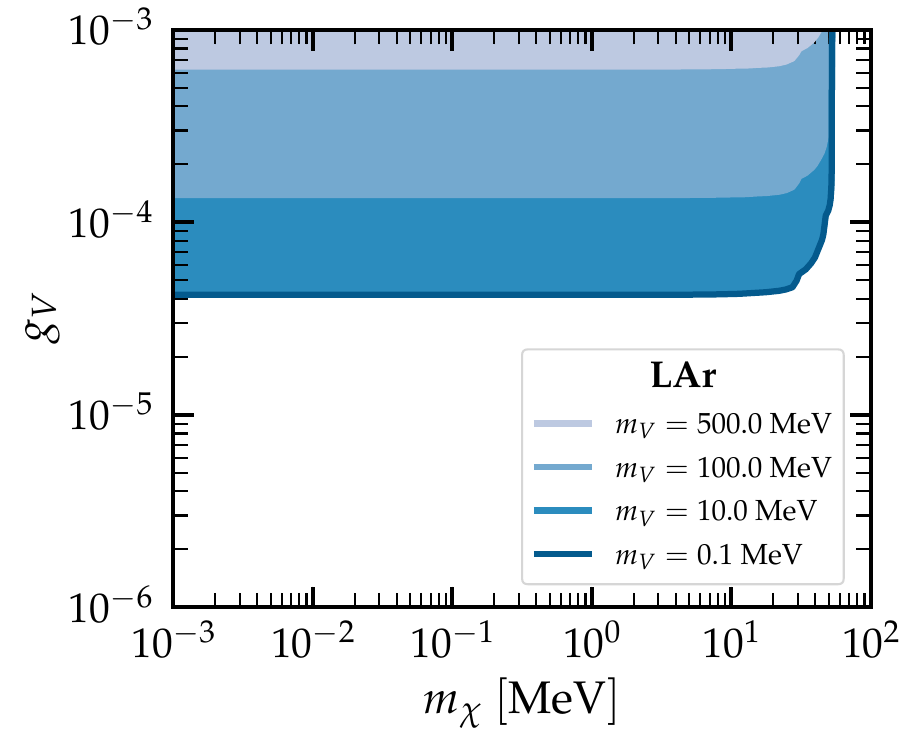}
\end{subfigure}

\begin{subfigure}{0.49\textwidth}
    \includegraphics[width=\textwidth]{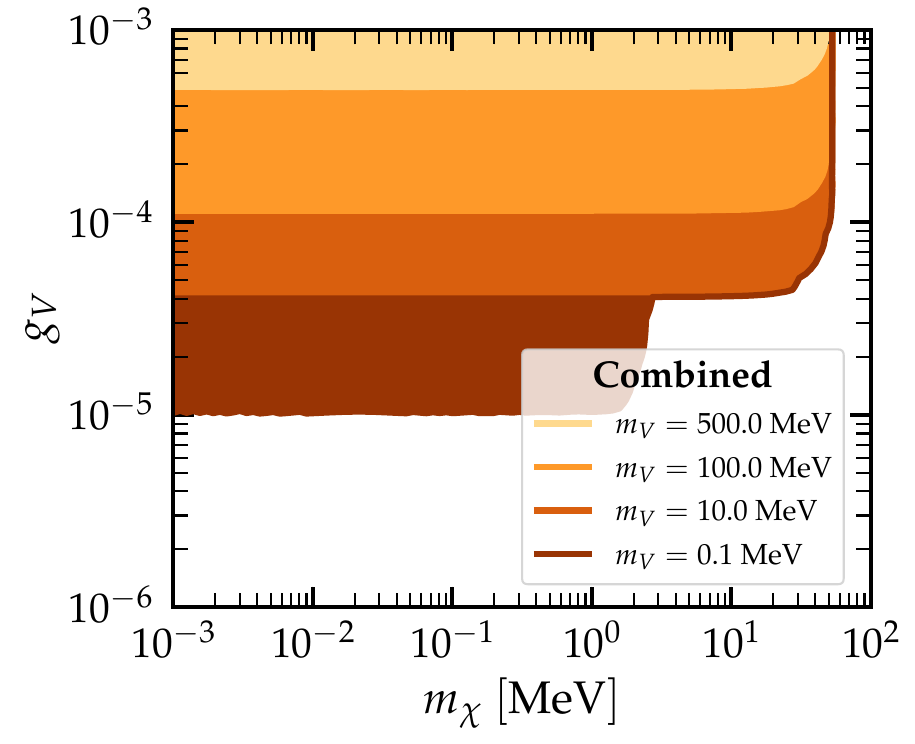}
\end{subfigure}
\hfill
\begin{subfigure}{0.49\textwidth}
    \includegraphics[width=\textwidth]{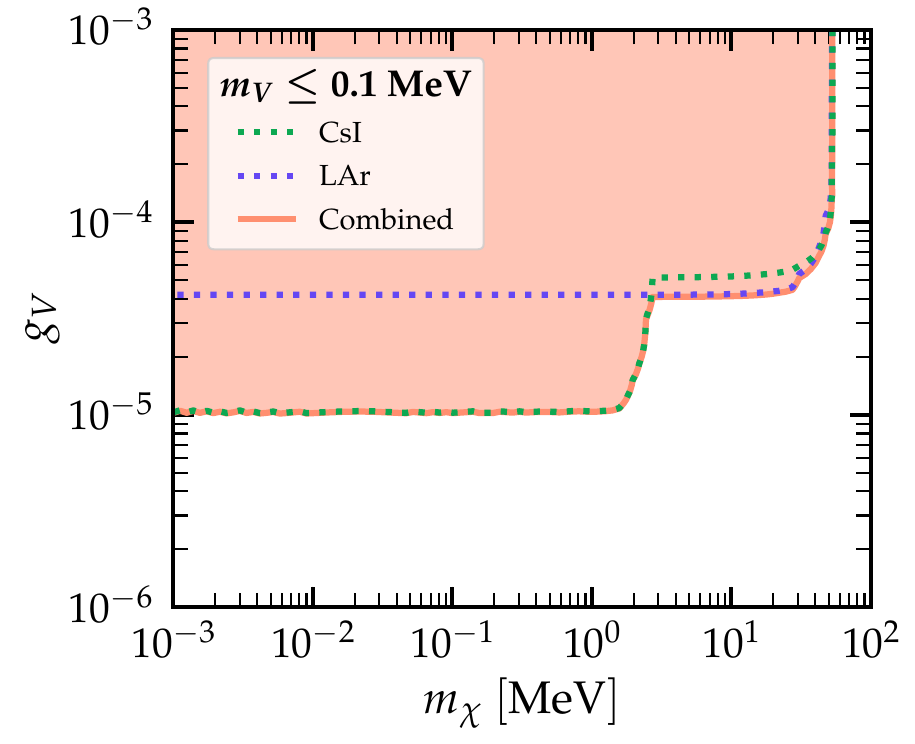}
\end{subfigure}
\caption{$90\%$ C.L. (2 d.o.f.) exclusion regions in the $m_\chi-g_V$ plane, for the vector-mediator case. Different fixed values of $m_V$ are considered.}
\label{fig:fixed-mv-vector-mediator}
\end{figure}

\begin{figure}
\centering
\begin{subfigure}{0.49\textwidth}
    \includegraphics[width=\textwidth]{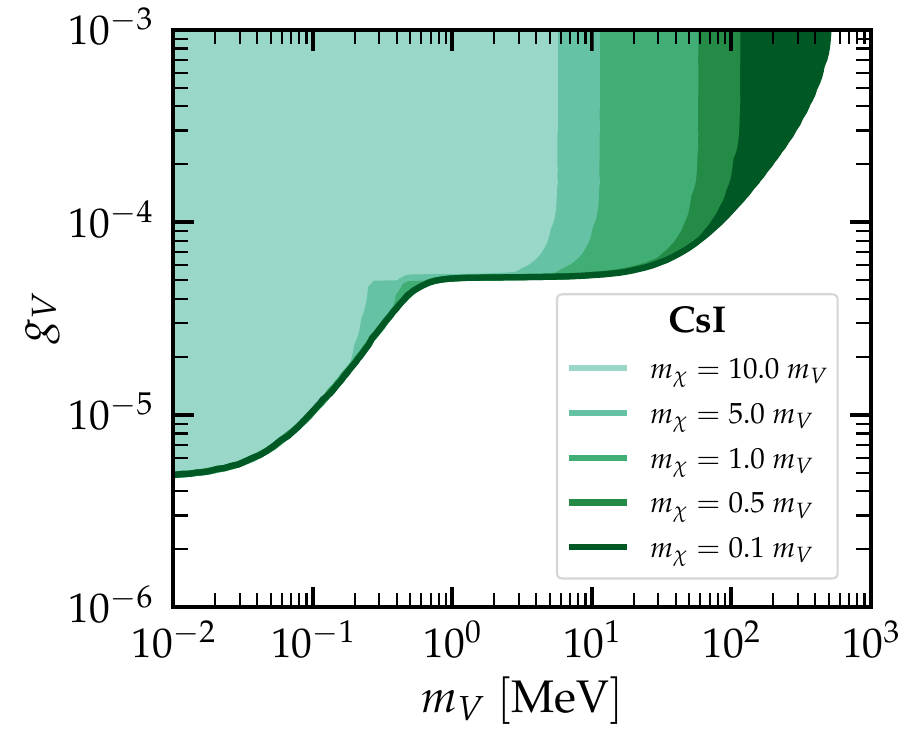}
\end{subfigure}
\hfill
\begin{subfigure}{0.49\textwidth}
    \includegraphics[width=\textwidth]{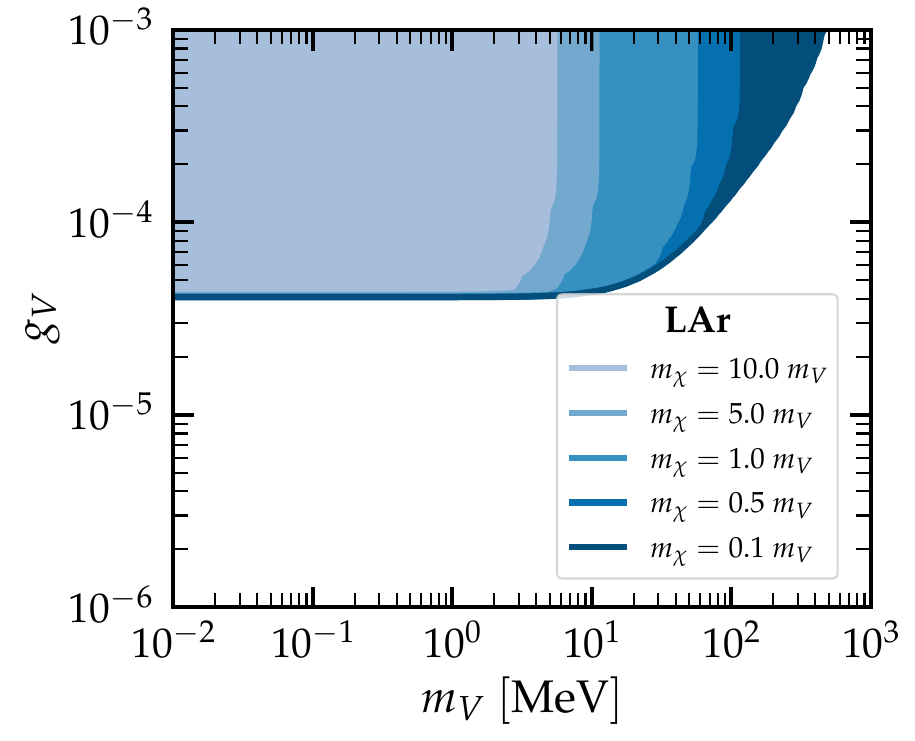}
\end{subfigure}

\begin{subfigure}{0.49\textwidth}
    \includegraphics[width=\textwidth]{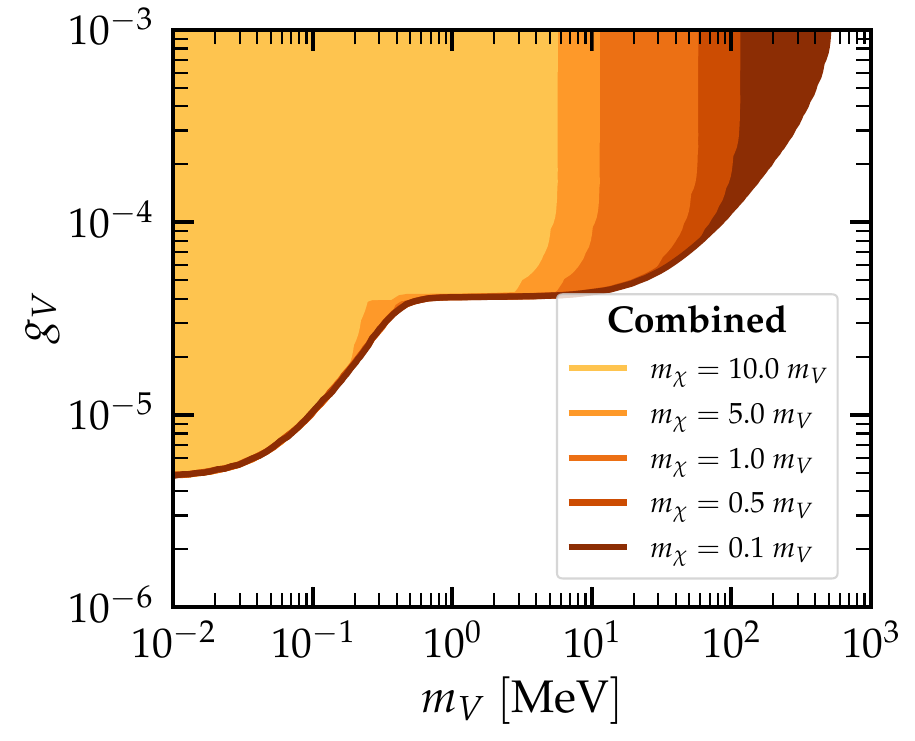}
\end{subfigure}
\hfill
\begin{subfigure}{0.49\textwidth}
    \includegraphics[width=\textwidth]{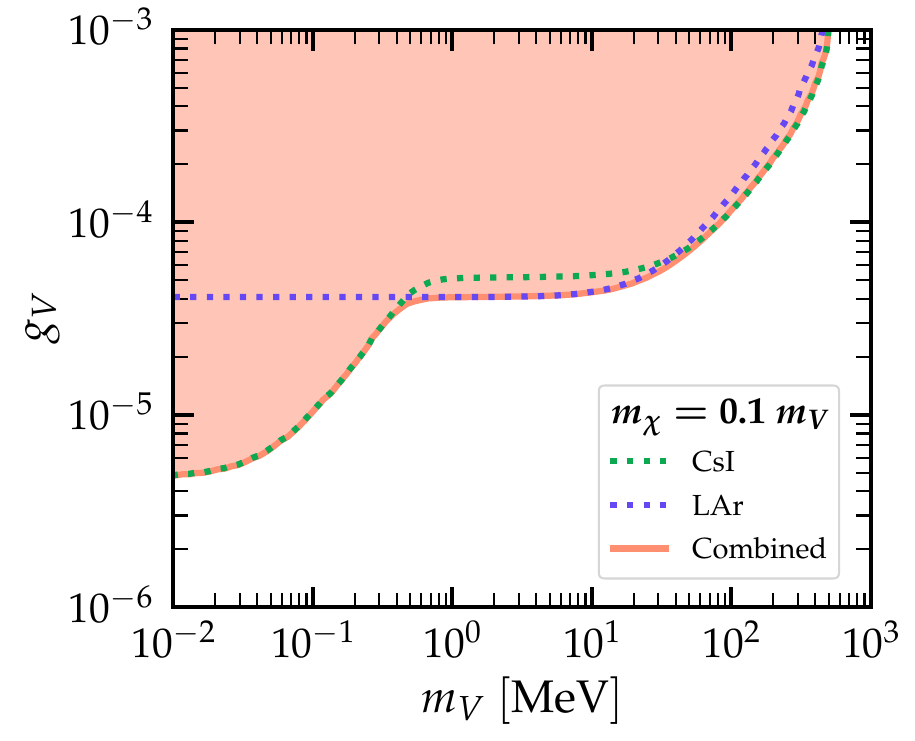}
\end{subfigure}
\caption{$90\%$ C.L. (2 d.o.f.) exclusion regions in the $m_V-g_V$ plane, for the vector-mediator case. Different values of $m_\chi$ proportional to $m_V$ are considered.} 
\label{fig:proportional-mchi-vs_mv-vector-mediator}
\end{figure}

\begin{figure}
\centering
\begin{subfigure}{0.49\textwidth}
    \includegraphics[width=\textwidth]{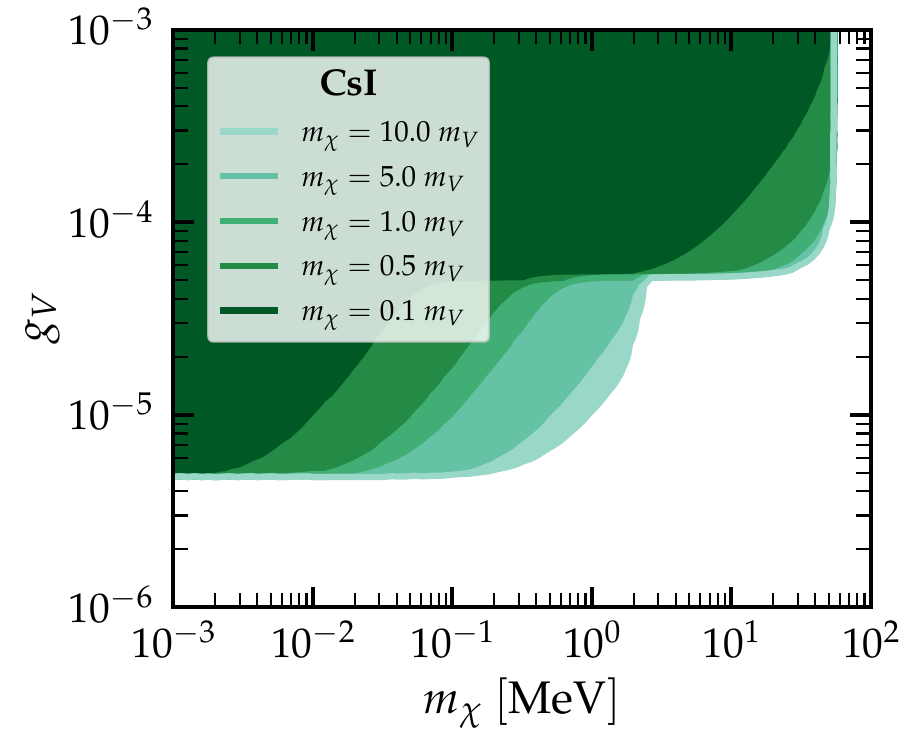}
\end{subfigure}
\hfill
\begin{subfigure}{0.49\textwidth}
    \includegraphics[width=\textwidth]{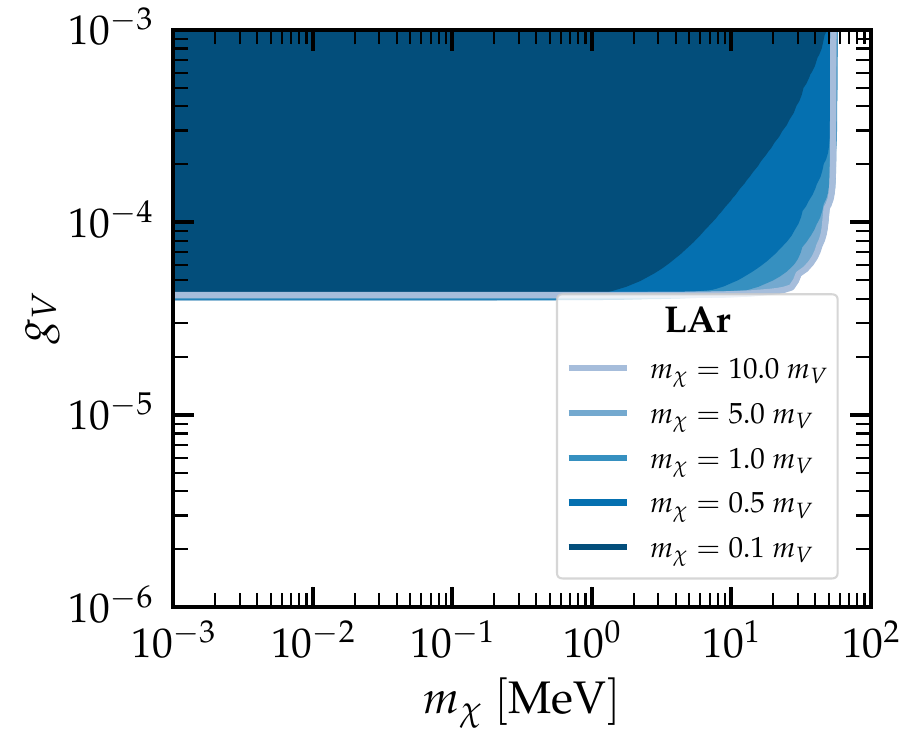}
\end{subfigure}

\begin{subfigure}{0.49\textwidth}
    \includegraphics[width=\textwidth]{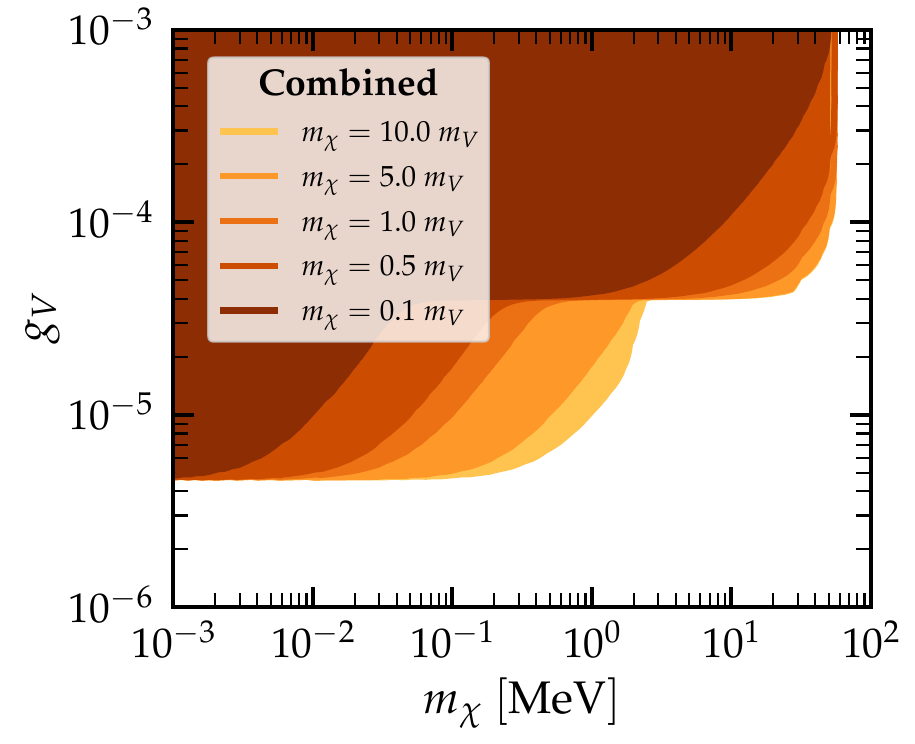}
\end{subfigure}
\hfill
\begin{subfigure}{0.49\textwidth}
    \includegraphics[width=\textwidth]{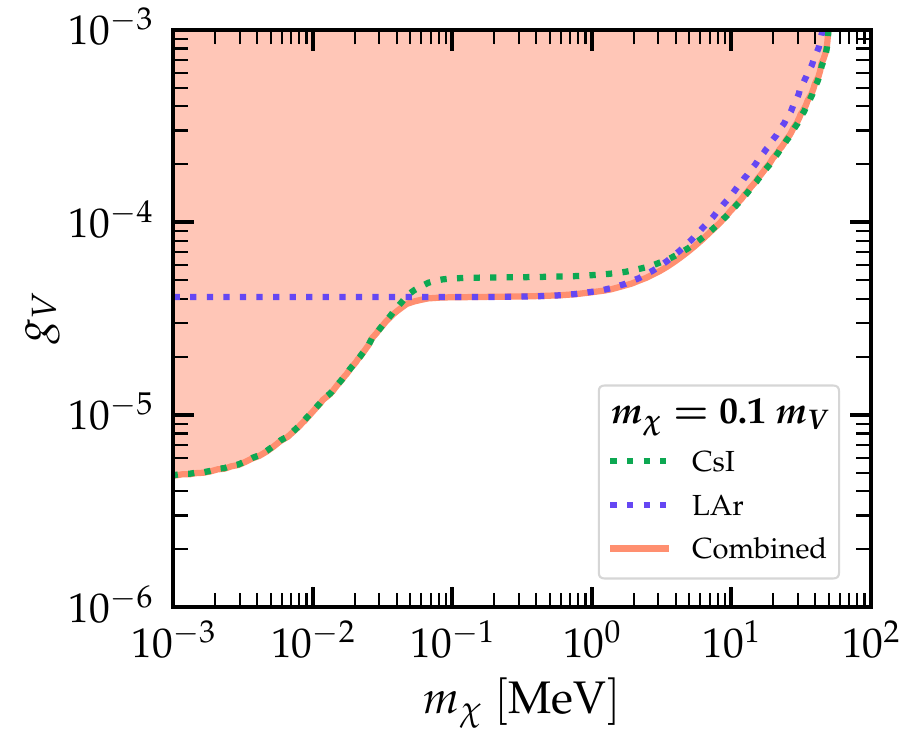}
\end{subfigure}
\caption{$90\%$ C.L. (2 d.o.f.) exclusion regions in the $m_\chi - g_V$ plane, for the vector-mediator case. Different values of $m_\chi$ proportional to $m_V$ are considered.} 
\label{fig:proportional-mchi-vs_mchi-vector-mediator}
\end{figure}

We start our analysis by fixing $m_\chi$ and letting the mediator mass and the coupling vary. In Fig.~\ref{fig:fixed-mchi-vector-mediator} we choose several fixed values of $m_\chi$ ($m_\chi = 0.1, 10$ and $50$ MeV, shown with darker to lighter shades) and plot the $90\%$ C.L. exclusion regions (assuming 2 d.o.f.) for the CsI (top-left) and LAr (top-right) COHERENT data.  We also perform a combined CsI+LAr data analysis (bottom-left) and we finally compare the three results (CsI, LAr and the combined CsI+LAr) for the most constraining case, corresponding to $m_\chi = 0.1~\mathrm{MeV}$ (bottom-right). The most notable difference among the two datasets is that CsI data lead to a significantly improved sensitivity compared to the LAr data in the low mass region, due to the inclusion of the ES events in the analysis. However, only the $m_\chi = 0.1~\mathrm{MeV}$ case seems to be sensitive to this effect. This is due to the fact that the mass of the produced DF is bounded from above. Indeed, as explained in Sec.~\ref{sec:prod}, there is an upper limit on  $m_\chi$ dictated by the kinematics of the up-scattering process. In particular, for \cevns ~on Cs or I nuclei,  $m_\chi \lesssim 53~\mathrm{MeV}$, while for ES the bound is lower, $m_\chi \lesssim 7~\mathrm{MeV}$.

In the intermediate region, LAr is more sensitive than CsI, whereas in the high mass region, CsI data lead to more stringent constraints. In the limit of heavy mediator masses~\footnote{This limit corresponds to the case of effective couplings $\epsilon^V \equiv \sqrt{2} g_V^2 / \left(G_F m_V^2 \right)$ explored in Ref.~\cite{Chen:2021uuw}.}, i.e. $m_V \gg |\mathbf{q}|$, or equivalently $m_V \gtrsim 100~\mathrm{MeV}$, the data start to lose sensitivity on $g_V$, since the cross section gets suppressed due to the presence of the mediator mass in the denominator. On the contrary, when $m_V \ll |\mathbf{q}|$ the cross section becomes independent of $m_V$ and the exclusion contour reaches a plateau. We further find that the $m_\chi = 50~\mathrm{MeV}$ benchmark leads to a less stringent constraint on $g_V$ in the low-mass region, compared to the $m_\chi = 10~\mathrm{MeV}$ one. This is due to the fact that the term proportional to $m_\chi$ in the kinetic factor of the cross section (see Eq.~\eqref{eq:xsec_DF_V}) is subtracting the term on the left. Thus, a larger coupling $g_V$ is required in order to produce a sizeable number of events, comparable to the experimental one. On the contrary, if $m_\chi$ decreases, then the coupling can also decrease and the corresponding constraint is stronger. Finally, in the limit $m_\chi \to 0$, the light mediator case from~\cite{DeRomeri:2022twg} is recovered, barring a small factor due to the absence of interference terms between the SM and the new physics interactions in the DF scenario. We have furthermore checked that the limits saturate for $m_\chi \lesssim 0.1$ MeV, at low $m_V$.

Next, we perform the same analysis, but fixing the mediator mass, $m_V$. Figure~\ref{fig:fixed-mv-vector-mediator} shows the corresponding result at 90\% C.L. in the $m_\chi-g_V$ plane for $m_V = 0.1,\,10,\,100$ and $500$ MeV, depicted with darker to lighter shades. In the CsI (top-left) panel, the impact of the ES events below the bound of $m_\chi \sim 7~\mathrm{MeV}$ is clearly visible. Notice however that the effect of the ES contribution only appears for  $m_V=0.1~\mathrm{MeV}$~\footnote{The case of $m_V = 1.0~\mathrm{MeV}$ was also computed, but it overlaps the $m_V=0.1~\mathrm{MeV}$ contour and it is not shown in the plot.}. For higher $m_V$ values the ES cross section given in Eq.~\eqref{eq:xsec_DF_ES_V} gets suppressed thus making its contribution to the total number of events negligible. Let us emphasize that, without ES, LAr data (top-right) provide better sensitivity on the coupling than CsI. For this reason, the combination of both datasets (bottom row) produces a stronger constraint. Notice also the kinematic limit which is again clearly visible in the figure, $m_\chi \sim 53~\mathrm{MeV}$ for every possible value of $m_V$.

In the last set of analyses we set $m_\chi$ proportional to $m_V$ and we let both the coupling and the mediator mass vary freely. In Fig.~\ref{fig:proportional-mchi-vs_mv-vector-mediator} we plot the $90 \%$ C.L. bounds on $g_V$ versus the mediator mass, for $m_\chi = 0.1,\,0.5,\,1,\,5,\,10~m_V$, from dark to light contours. The main difference with respect to the previous figures is that the different contours are not shifted vertically but horizontally. Take, for example, the case with $m_\chi = 10~m_V$. When $m_V = 5~\mathrm{MeV}$, $m_\chi$ is close to its upper bound and for values higher than that, the up-scattering production of the DF is kinematically forbidden, hence only the SM \cevns ~remains. If we focus now on the contour for $m_\chi = 0.5~\mathrm{MeV}$, the upper bound on $m_\chi$ is achieved at higher values of $m_V$, hence it is more constraining.

Finally, Fig.~\ref{fig:proportional-mchi-vs_mchi-vector-mediator} depicts the same data as Fig.~\ref{fig:proportional-mchi-vs_mv-vector-mediator}, except that it is plotted with respect to the DF mass. The upper bound of $m_\chi \sim 53~\mathrm{MeV}$ is the same for every contour line. In the CsI panel (top-left) the ES contribution is shifted to the left when $m_\chi$ gets smaller with respect to $m_V$. The reason is that when the ES contribution is allowed ($m_\chi \lesssim 7~\mathrm{MeV}$), the mediator mass is large and the cross section gets suppressed.

\subsection{Scalar mediator}
Similarly to the vector-mediator discussion, in this subsection we refer to the Lagrangian given in Eq.~\eqref{eq:Lagr} and to the cross sections given in Eqs.~\eqref{eq:xsec_DF_S} and \eqref{eq:xsec_DF_ES_S}. Thanks to the redefinition of the coupling $g_S \equiv \sqrt{g_{\chi_L} g_f}$, with $f = u,\,d$ for \cevns~and $f=e^-$ for ES, we can express our results in terms of the relevant parameters, namely the DF mass ($m_\chi$), the mediator mass ($m_S$) and one coupling ($g_S$). In analogy to the vector-mediator case described above, this simplification is based on the assumption that the scalar mediator has universal couplings to both quarks and electrons.

\begin{figure}
\centering
\begin{subfigure}{0.49\textwidth}
    \includegraphics[width=\textwidth]{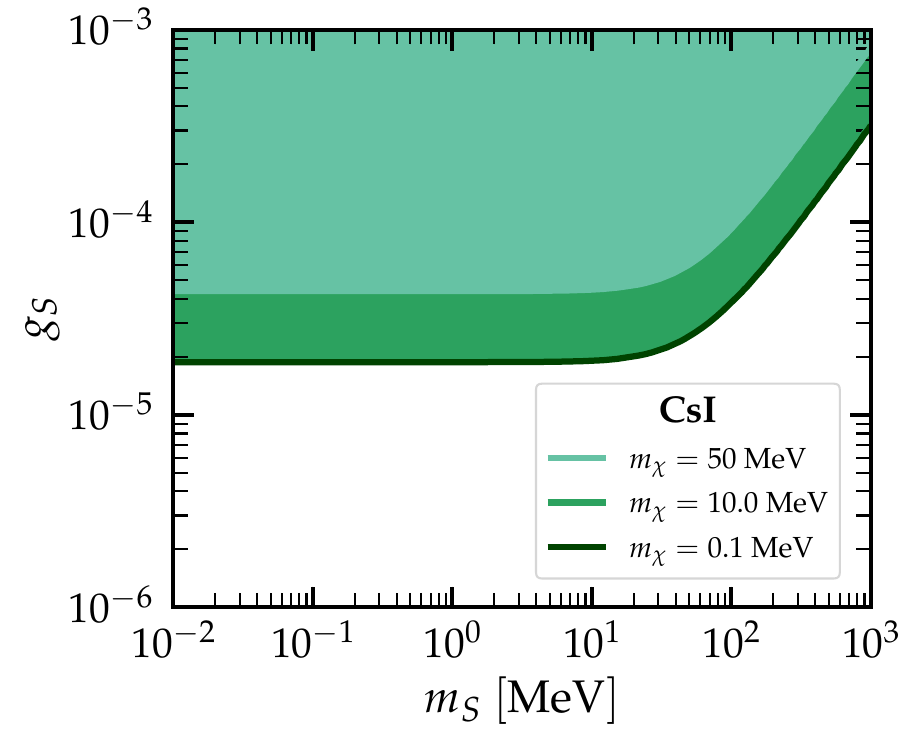}
\end{subfigure}
\hfill
\begin{subfigure}{0.49\textwidth}
    \includegraphics[width=\textwidth]{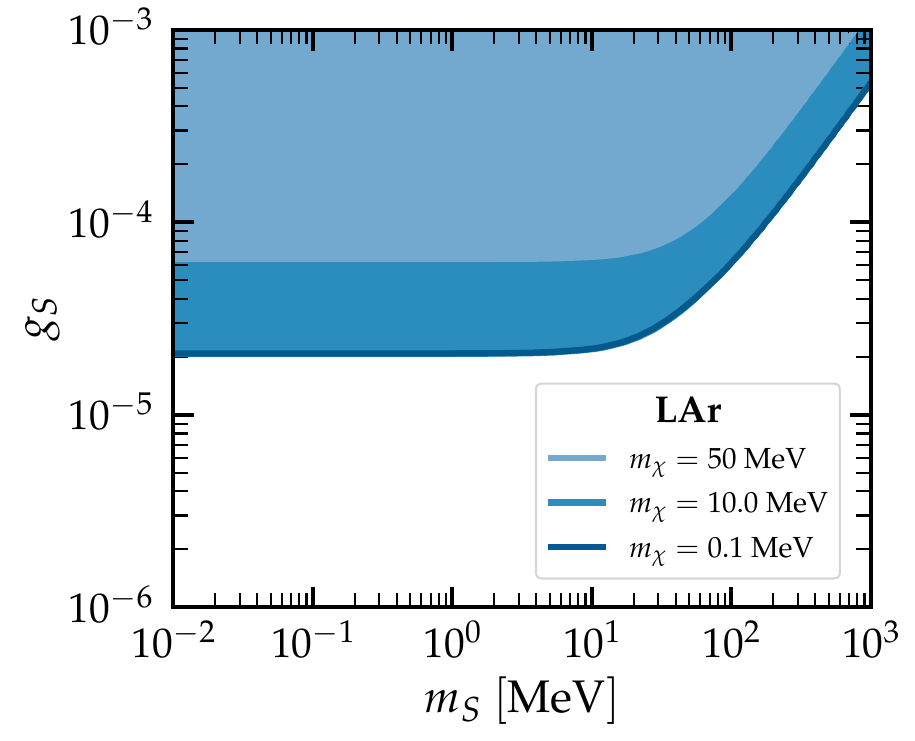}
\end{subfigure}

\begin{subfigure}{0.49\textwidth}
    \includegraphics[width=\textwidth]{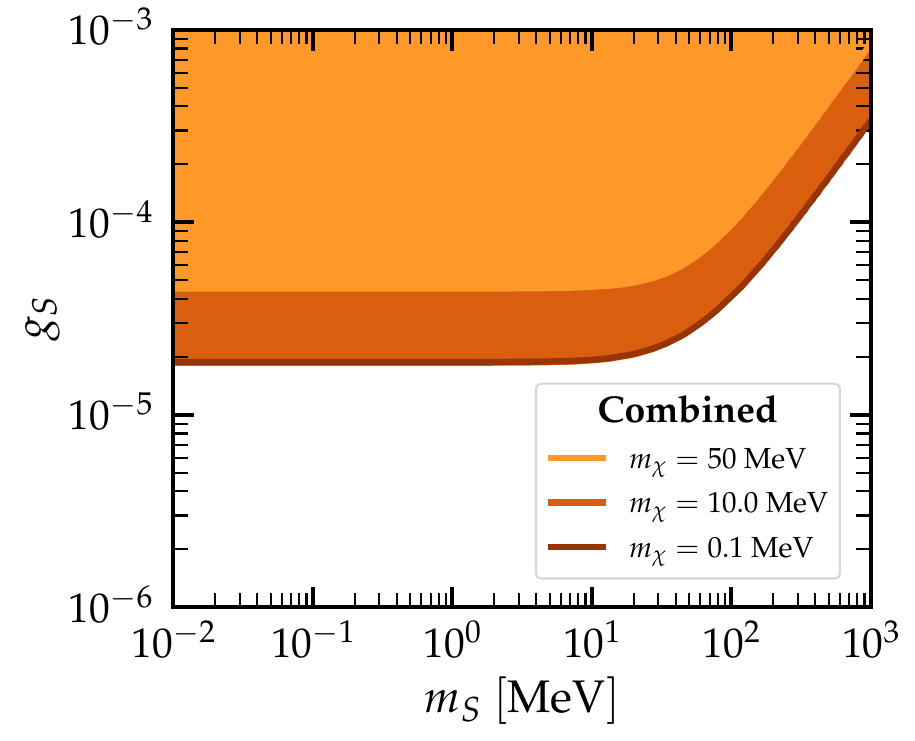}
\end{subfigure}
\hfill
\begin{subfigure}{0.49\textwidth}
    \includegraphics[width=\textwidth]{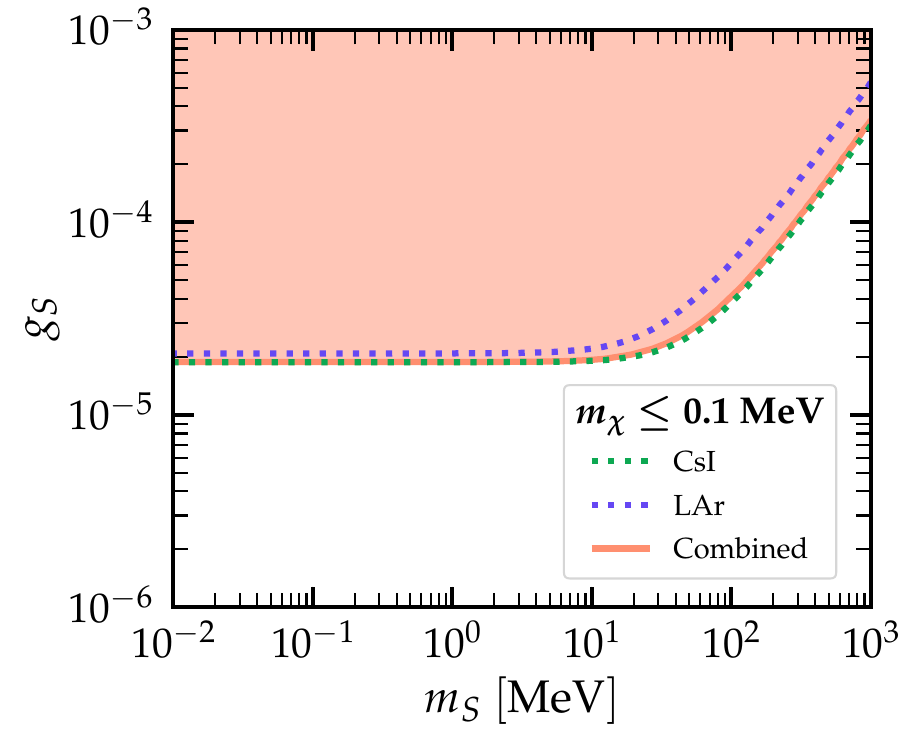}
\end{subfigure}
\caption{$90\%$ C.L. (2 d.o.f.) exclusion regions for the scalar-mediator case, in the $m_S - g_S$ plane. Different fixed values of $m_\chi$ are considered.} 
\label{fig:fixed-mchi-scalar-mediator}
\end{figure}

\begin{figure}
\centering
\begin{subfigure}{0.49\textwidth}
    \includegraphics[width=\textwidth]{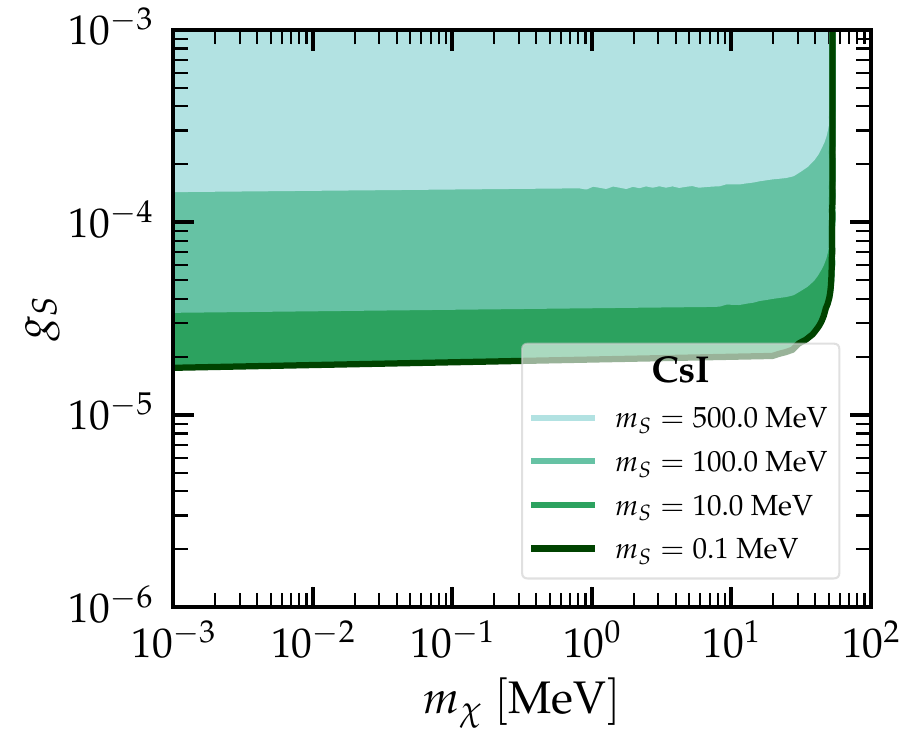}
\end{subfigure}
\hfill
\begin{subfigure}{0.49\textwidth}
    \includegraphics[width=\textwidth]{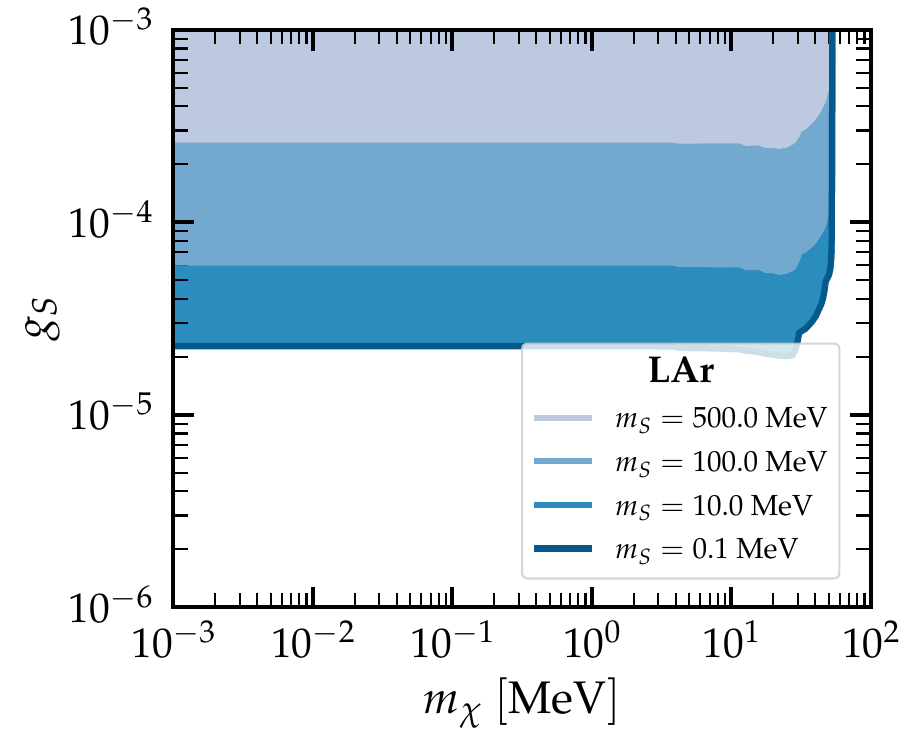}
\end{subfigure}

\begin{subfigure}{0.49\textwidth}
    \includegraphics[width=\textwidth]{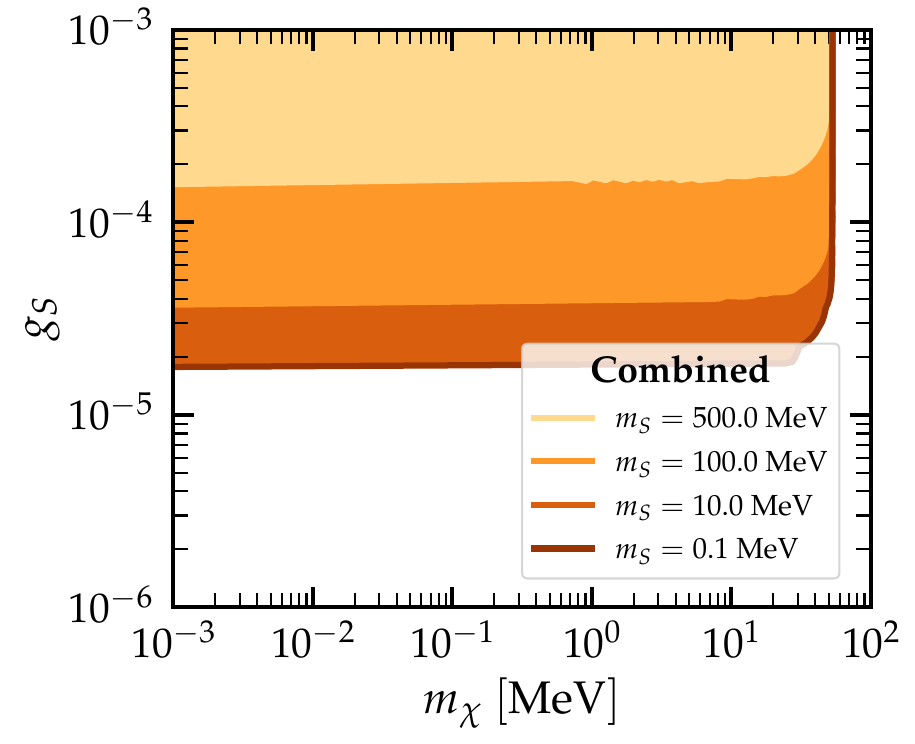}
\end{subfigure}
\hfill
\begin{subfigure}{0.49\textwidth}
    \includegraphics[width=\textwidth]{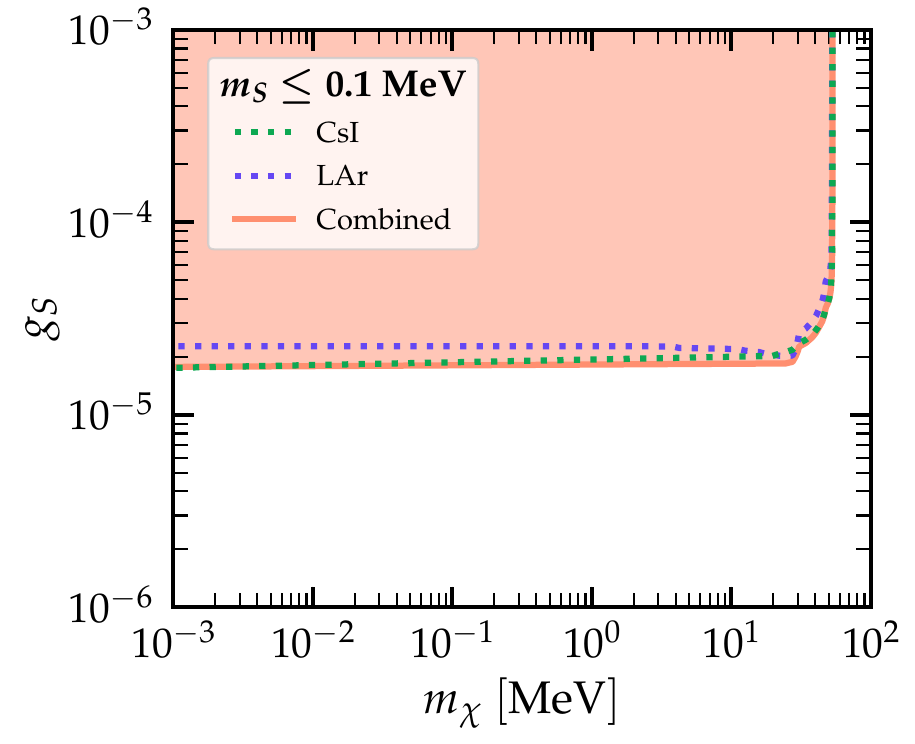}
\end{subfigure}
\caption{$90\%$ C.L. (2 d.o.f.) exclusion regions for the scalar mediator case, in the $m_\chi - g_S$ plane. Different fixed values of $m_S$ are considered.} 
\label{fig:fixed-ms-scalar-mediator}
\end{figure}

\begin{figure}
\centering
\begin{subfigure}{0.49\textwidth}
    \includegraphics[width=\textwidth]{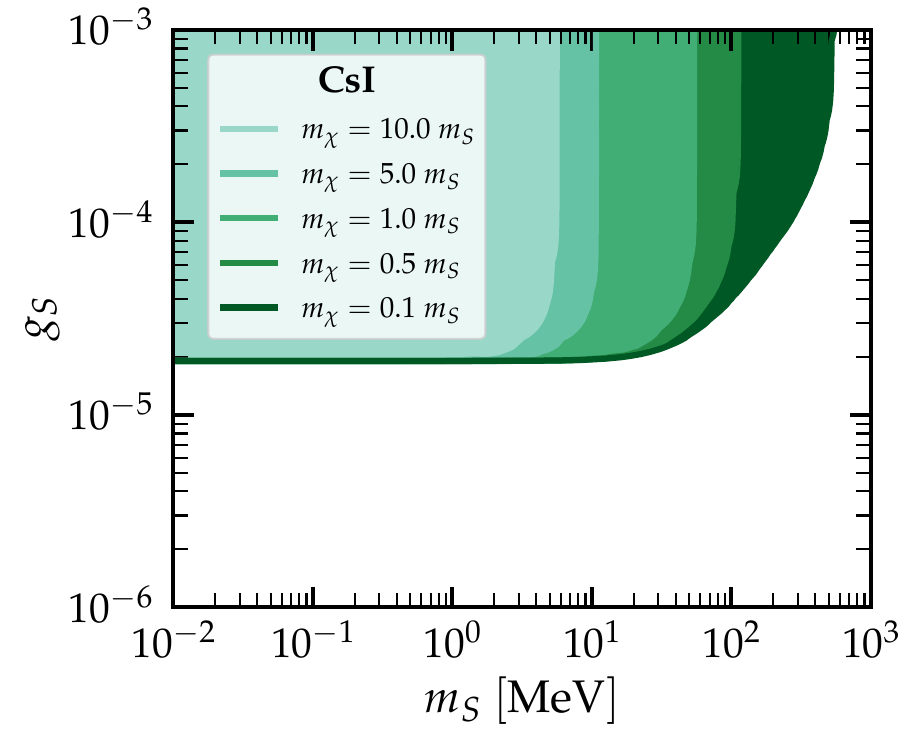}
\end{subfigure}
\hfill
\begin{subfigure}{0.49\textwidth}
    \includegraphics[width=\textwidth]{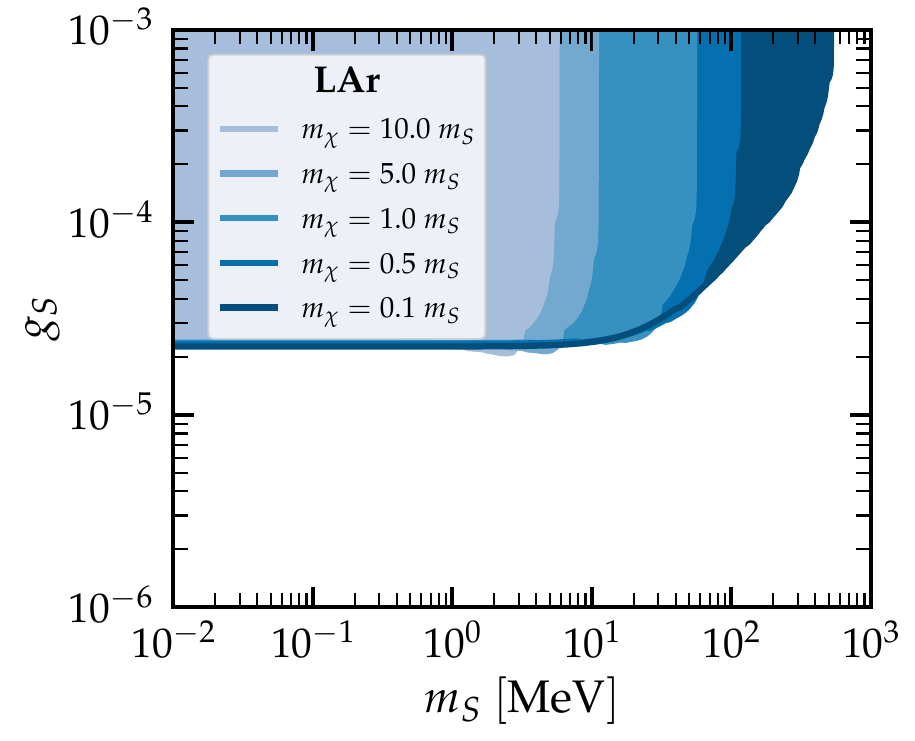}
\end{subfigure}

\begin{subfigure}{0.49\textwidth}
    \includegraphics[width=\textwidth]{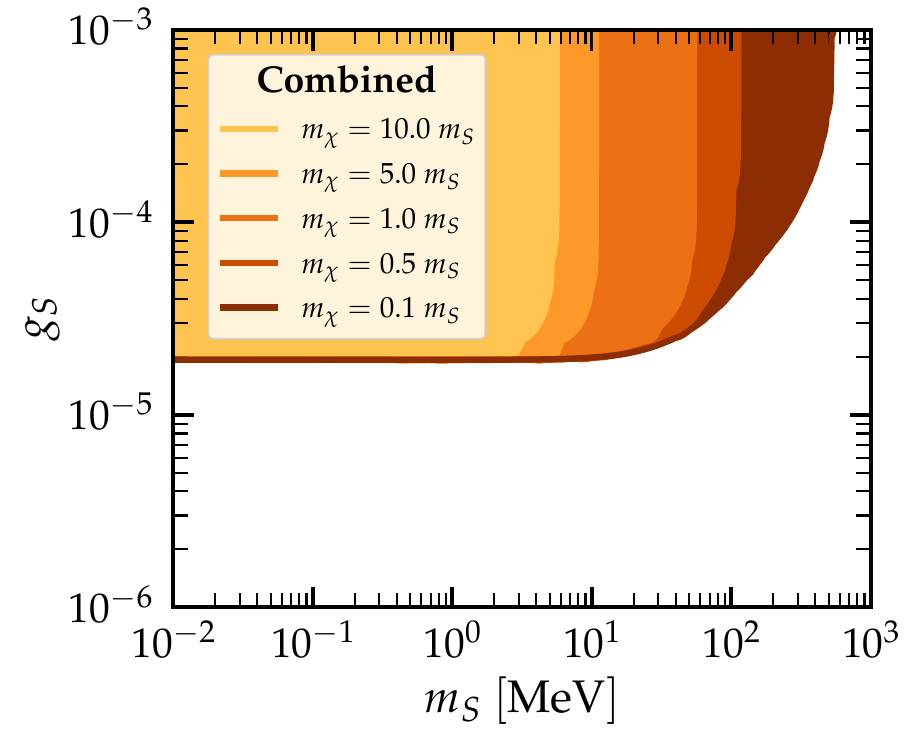}
\end{subfigure}
\hfill
\begin{subfigure}{0.49\textwidth}
    \includegraphics[width=\textwidth]{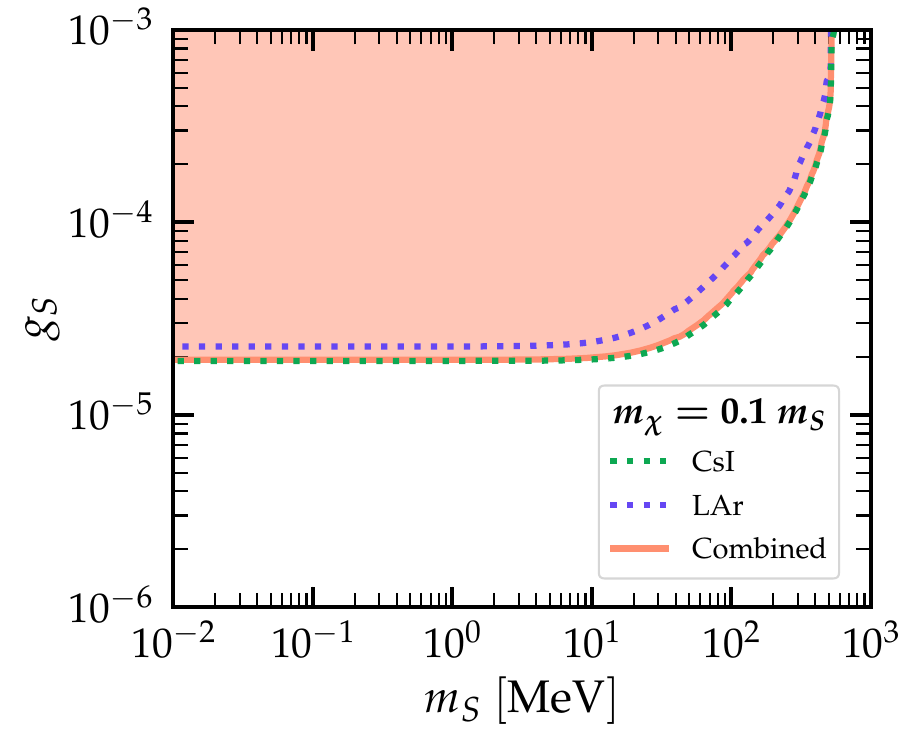}
\end{subfigure}
\caption{$90\%$ C.L. (2 d.o.f.) exclusion regions for the scalar mediator case, in the $m_S - g_S$ plane. Different values of $m_\chi$ proportional to $m_S$ are considered.} 
\label{fig:proportional-mchi-vs_ms-scalar-mediator}
\end{figure}

\begin{figure}
\centering
\begin{subfigure}{0.49\textwidth}
    \includegraphics[width=\textwidth]{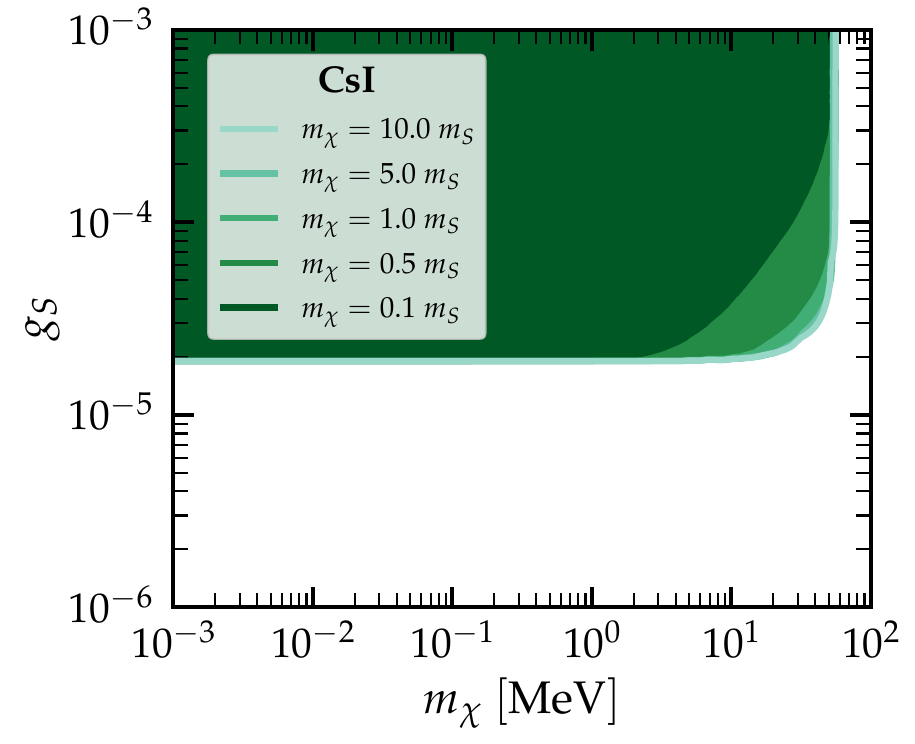}
\end{subfigure}
\hfill
\begin{subfigure}{0.49\textwidth}
    \includegraphics[width=\textwidth]{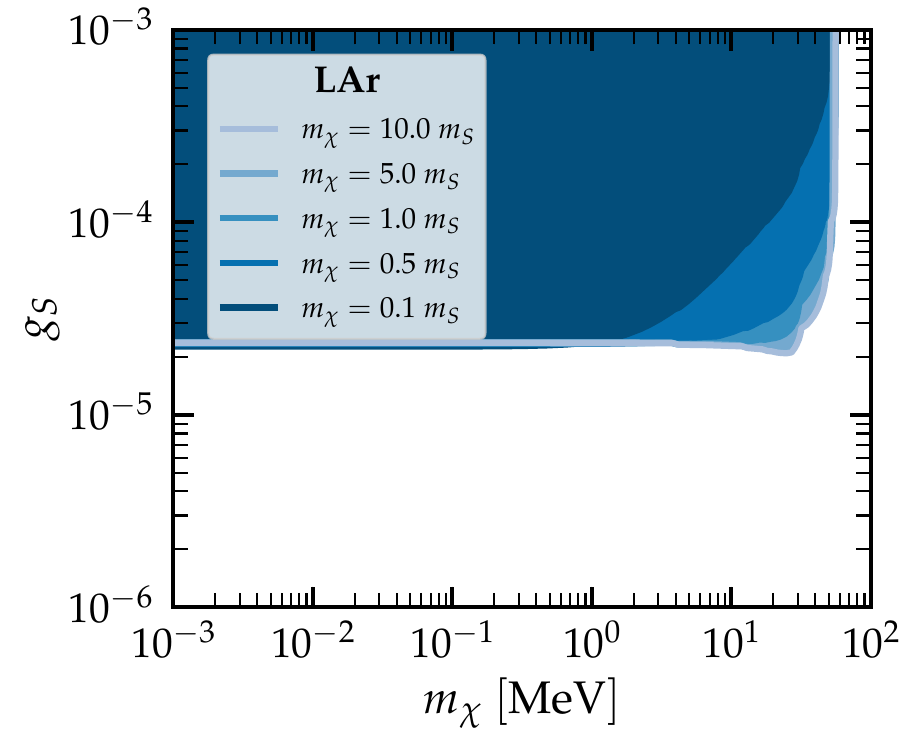}
\end{subfigure}

\begin{subfigure}{0.49\textwidth}
    \includegraphics[width=\textwidth]{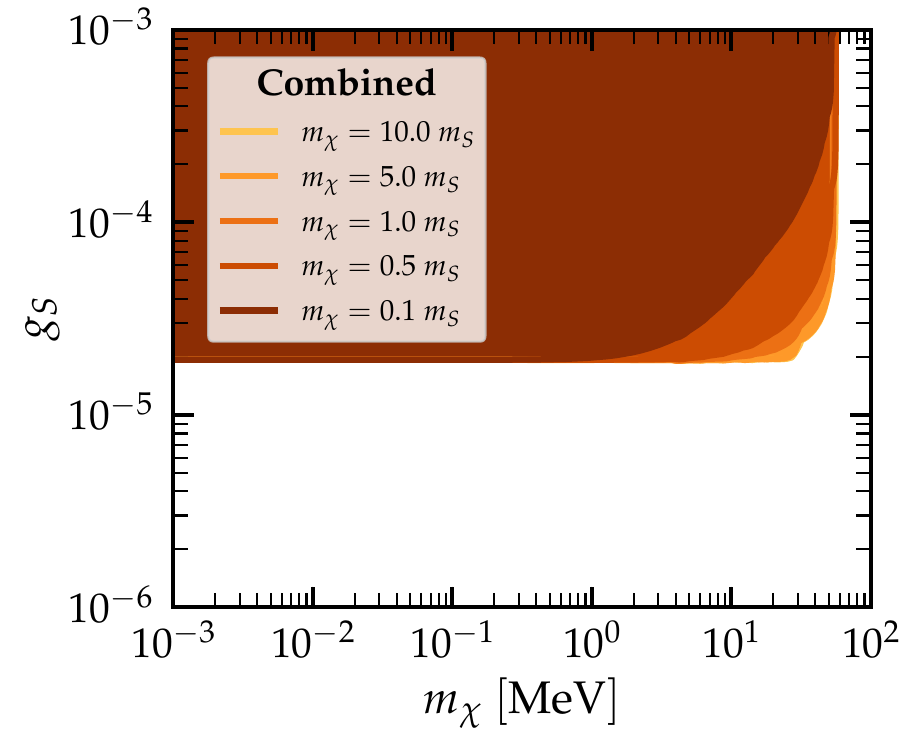}
\end{subfigure}
\hfill
\begin{subfigure}{0.49\textwidth}
    \includegraphics[width=\textwidth]{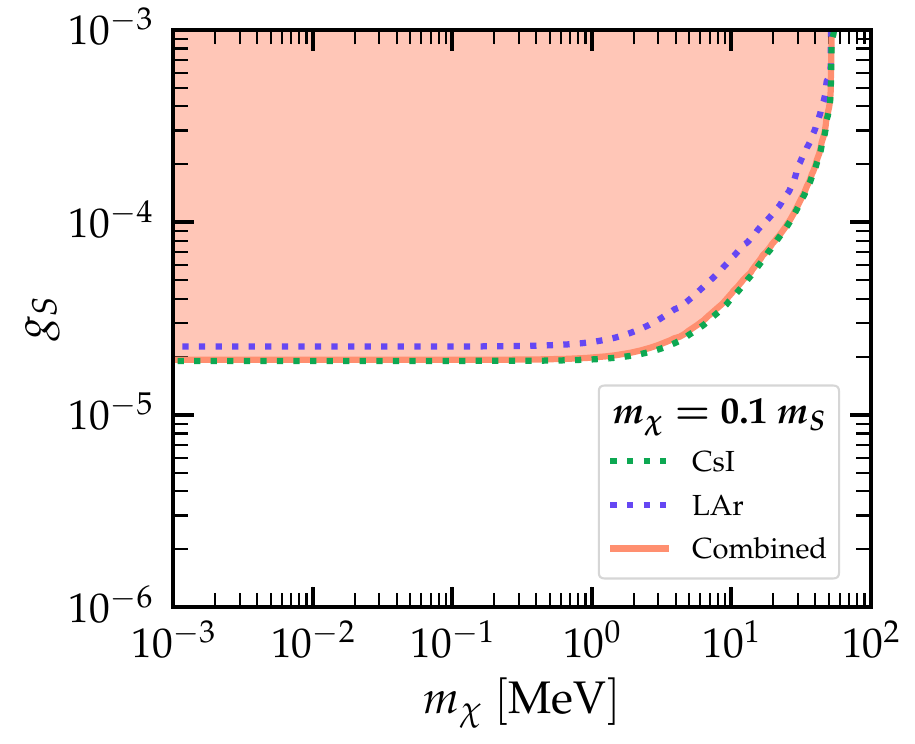}
\end{subfigure}
\caption{$90\%$ C.L. (2 d.o.f.) exclusion regions for the scalar mediator case, in the $m_\chi - g_S$ plane. Different values of $m_\chi$ proportional to $m_S$ are considered. (With respect to Fig.~\ref{fig:proportional-mchi-vs_ms-scalar-mediator} only the x-axis is changed.)} 
\label{fig:proportional-mchi-vs_mchi-scalar-mediator}
\end{figure}

In Figs.~\ref{fig:fixed-mchi-scalar-mediator},~\ref{fig:fixed-ms-scalar-mediator},~\ref{fig:proportional-mchi-vs_ms-scalar-mediator} and~\ref{fig:proportional-mchi-vs_mchi-scalar-mediator} we present the $90\%$ C.L. (2 d.o.f.) exclusion regions for the same benchmarks described in the previous subsection. Because of the similarities in the general interpretation of the two results, we refer the reader to the vector mediator discussion in  Sec.~\ref{subsec:results-vector-mediator}, while here we will only comment about the main differences between scalar and vector scenarios.

The first relevant variation is that the ES contribution ---affecting the CsI data analysis--- is negligible in the scalar-mediator case. Notice, for instance, that  in the corresponding constraints obtained in Figs.~\ref{fig:fixed-mchi-vector-mediator} and~\ref{fig:fixed-mv-vector-mediator}, the ES contribution led to a ``bump" in the exclusion regions at $m_\chi \lesssim 7$ MeV, that is now absent.  This is due to the different expressions for the ES cross sections given in Eqs. \eqref{eq:xsec_DF_ES_V} and \eqref{eq:xsec_DF_ES_S}. The scalar cross section is suppressed
by a factor $m_e E_{\mathrm{er}}/E_\nu^2$
compared to the leading terms of the vector case. Indeed, in the ES cross section for the vector mediator there is a constant term that is always present regardless of the values of the masses or the energies appearing in the kinematic terms.
Without the ES contribution, the inferred bounds on the coupling $g_S$ saturate already at $m_\chi \lesssim 10~\mathrm{MeV}$ in Fig.~\ref{fig:fixed-mchi-scalar-mediator} and at $m_S \lesssim 10~\mathrm{MeV}$ in Fig.~\ref{fig:fixed-ms-scalar-mediator}. 

At last, let us comment about how our results compare to those of Ref.~\cite{Brdar:2018qqj}. The bounds obtained in that reference rely on some assumptions, for instance $F_W^2 = 1$. Taking into account differences in the definition of the relevant coupling ($\bar{y}_s$ vs $g_S$), we notice that our results are in general comparable: while we include a more recent dataset with more statistics, we also perform a more sophisticated statistical analysis.

\subsection{Comparison to other constraints}
\begin{figure}
\centering
\begin{subfigure}{0.49\textwidth}
    \includegraphics[width=\textwidth]{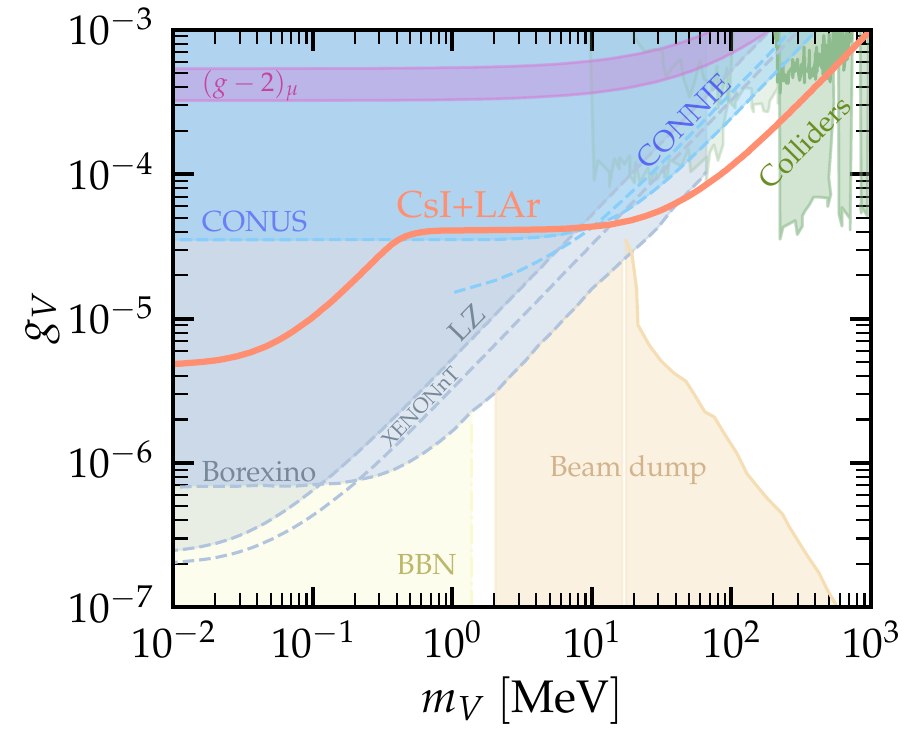}
\end{subfigure}
\hfill
\begin{subfigure}{0.49\textwidth}
    \includegraphics[width=\textwidth]{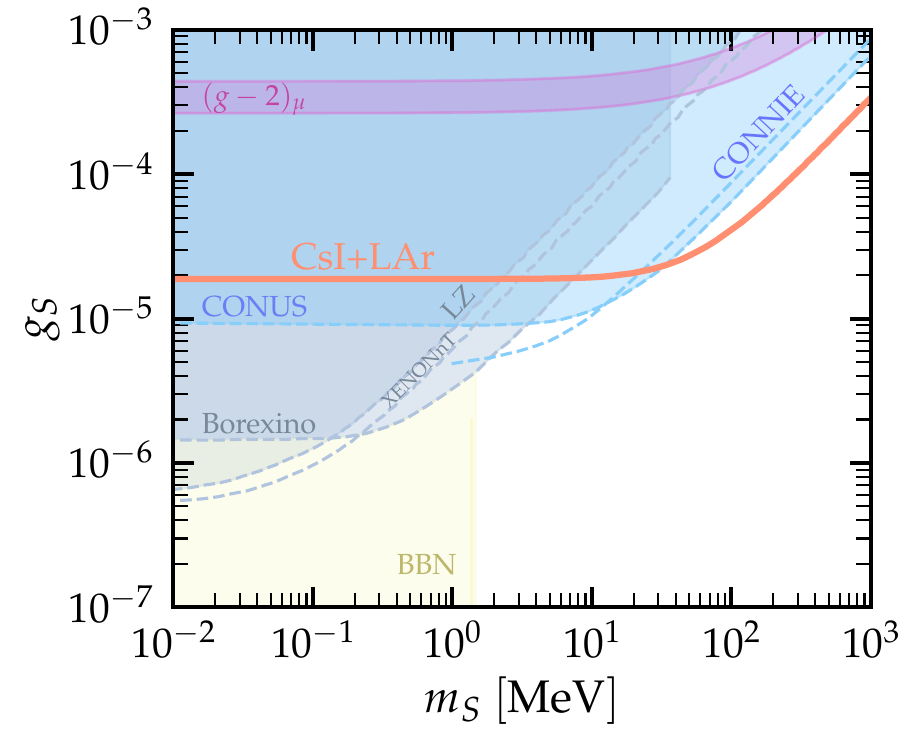}
\end{subfigure}
\caption{$90\%$ C.L. (2 d.o.f.) exclusion regions for the vector- (left) and scalar-mediator (right) case. 
The orange contours are obtained from the combined analysis of CsI+LAr data, and assuming a massless DF. We refer to the main text for more details regarding this approximation.}
\label{fig:fixed-mchi-detectors-comparison}
\end{figure}

In this subsection we discuss the limits obtained in this work in comparison with other existing bounds in the literature. In principle, limits from other experiments should be carefully recast in terms of the up-scattering scenario that we considered. However we recall here that our exclusion contours saturate for $m_\chi \lesssim 0.1$ MeV, and eventually coincide with the limits expected for  $m_\chi \to 0$. Given these considerations, we can compare to existing constraints on \cevns~or neutrino ES with light vector mediators.
In Fig.~\ref{fig:fixed-mchi-detectors-comparison} we hence show, as for comparison, our result from the combined analysis together with limits from other experimental searches on light mediators (vector on the left and scalar on the right). We display limits from other \cevns~experiments, in particular CONNIE~\cite{CONNIE:2019xid}, CONUS~\cite{CONUS:2021dwh} and Dresden-II~\cite{AristizabalSierra:2022axl}, from collider experiments~\cite{AtzoriCorona:2022moj}, like LHCb~\cite{LHCb:2017trq} and BaBar~\cite{BaBar:2014zli}, from rare meson decays at NA48~\cite{NA482:2015wmo}, from Borexino solar neutrino data~\cite{Coloma:2022umy} and from the analyses~\cite{A:2022acy} of multi-ton DM experiments (XENONnT~\cite{XENONCollaboration:2022kmb} and LZ~\cite{LZ:2022ufs}). Let us finally comment that limits from deep inelastic scattering (CHARM-II), from reactor neutrinos (TEXONO) and from solar neutrinos (BOREXINO) have been set on effective interactions leading to the up-scattering production of a DF~\cite{Chen:2021uuw}. While a direct comparison with these results is not possible as we are considering light mediators, we can notice that COHERENT can explore DF masses above the reach of TEXONO and BOREXINO, and a bit below CHARM-II. To conclude, we see that the limits here obtained using COHERENT data improve upon existing ones for mediators with masses in the range $70~\mathrm{MeV} \lesssim m_V \lesssim 110~\mathrm{MeV}$ and $m_S \gtrsim 40$ MeV.

\subsection{The dark fermion as the dark matter}
The DF produced through neutrino up-scattering at COHERENT, belonging to a dark sector, could also constitute part or the totality of the DM observed in the Universe. The evaluation of the cosmic abundance of $\chi$ will depend on all possible production mechanisms. In this respect, in addition to those induced by the interactions in Eq.~\eqref{eq:Lagr}, new ones could arise in UV-completions of our phenomenological model. 
While a rigorous discussion about the viability of $\chi$ as the DM would require a dedicated analysis and it is therefore beyond the scope of this work, in the following, we will make few considerations about its stability, focusing on the vector-mediator scenario. Similar considerations can also be made for the scalar-mediator case. We refer the reader to Refs.~\cite{Brdar:2018qqj,Hurtado:2020vlj,Chen:2021uuw,Li:2020pfy} for further discussions.

First of all we can note that, given the interactions allowed by the Lagrangian in Eq.~\eqref{eq:Lagr}, there are only two possibilities for $\chi$ to decay, at tree-level. If $m_\chi > m_V$, $\chi$ can decay into $ \chi \to V  \nu_\alpha$.  If instead $m_\chi < m_V$, the three-body decay $ \chi \to \nu_\alpha  f \bar{f}$ is allowed, via an off-shell mediator. Given the interactions in Eq.~\eqref{eq:Lagr} and the $\chi$ mass range accessible at COHERENT, eventually only the channel $\chi \to \nu_\alpha e^+ e^-$ is kinematically allowed. 
In principle, the tree-level decay $\chi \to \nu_\alpha \nu_\beta \bar{\nu}_\beta$ could also be allowed, depending on the nature of the interactions given in Eq.~\eqref{eq:Lagr}. However, if the new mediator $V$ couples only to charged fermions (e.g. in a dark-photon scenario, see for example Ref.~\cite{Jho:2020jfz}), the tree-level decay into three neutrinos is forbidden.  We assume this possibility and we hence estimate the tree-level decay widths as following 
\begin{align}
    \Gamma(\chi \to V \nu_\alpha) &= \dfrac{g_{\chi_L}^2}{32\pi} m_\chi \left(1-\dfrac{m_V^2}{m_\chi^2}\right) \left(1 + \dfrac{m_\chi^2}{m_V^2}-2\dfrac{m_V^2}{m_\chi^2}\right), \nonumber \\[4pt]
    \Gamma(\chi \to \nu_\alpha e^- e^+) &= \dfrac{g_V^4}{768 \pi^3}\, m_\chi^5 \, I(m_\chi, m_V, \mu_2)\, , \label{eq:chi-3body-decay}
\end{align}
where $I(m_\chi, m_V, \mu_2)$ reads
\begin{align}
    I(m_\chi, m_V, \mu_2) \equiv \int_{0}^{1-4\mu_2} \mathrm{d}x_1 \, &\dfrac{\lambda^{1/2}(1-x_1, \mu_2, \mu_2)}{\left[m_\chi^2(1-x_1) - m_V^2\right]^2}\dfrac{x_1^2}{(1-x_1)^3} \bigg\{6(1+2\mu_2) \bigg. \nonumber \\[4pt]
    &\bigg.- x_1\left[15 + \lambda(1-x_1, \mu_2, \mu_2) + 24\mu_2 + 3x_1^2 - 12x_1(1+\mu_2)\right]\bigg\}.
\end{align}
Here $\mu_2 \equiv m_e^2 / m_\chi^2$ and $\lambda(x,y,z)=x^2+y^2+z^2-2xy-2xz-2yz$ is the K\"all\'en function. Equation~\eqref{eq:chi-3body-decay} is valid in both the heavy- and light-mediator regimes. Let us comment that to compare with other results in the literature, for instance~\cite{Bertuzzo:2018itn,Jho:2020jfz,Jodlowski:2020vhr}, where the heavy-mediator approximation is assumed ($m_V \gg m_\chi$),  one may use $2 I(m_\chi, m_V, \mu_2) \approx [I_2(0,\sqrt{\mu_2},\sqrt{\mu_2}) + 2I_1(0,\sqrt{\mu_2},\sqrt{\mu_2})]/m_V^4$. The kinematical functions $I_1$ and $I_2$ are defined, e.g., in Ref.~\cite{Helo:2010cw}. From these expressions, we can notice that for $m_\chi < 2 m_e$ no tree-level decays are kinematically allowed. Radiative decays such as $\chi \to \nu \gamma$, $\chi \to \nu \gamma \gamma \gamma$ or $\chi \to \nu \nu \nu$ could be, on the other hand, open. These decays have been computed assuming effective operators in~\cite{Dror:2020czw}. \\
Based on the previous decay rates, we estimate the lifetime of $\chi$ and compare it to the age of the Universe. 
We find that when assuming a very light mediator ($m_V \sim 1$ keV) and values of couplings testable at COHERENT, the tree-level decays given in Eq.~\eqref{eq:chi-3body-decay} are efficient enough to make $\chi$ decay before $t\sim 1$ sec, independently of its mass. In this case, $\chi$ could not be the DM, but also would not affect the BBN as it would decay at earlier epochs. When the mediator is heavy ($m_V \gtrsim 1$ MeV), $\chi$ would still decay faster than the age of the Universe, in the region of parameters where the decay channels in Eq.~\eqref{eq:chi-3body-decay} are kinematically allowed. Finally, should the mediator be heavy ($m_V \gtrsim 1$ MeV) and the DF light enough ($m_\chi < 1$ MeV) then the tree-level decays would not be kinematically accessible. Estimations of radiative decays~\cite{Dror:2020czw,PhysRevLett.124.181301,Li:2020pfy} also indicate that they would not be very efficient thus making $\chi$ a long-lived particle and possibly a viable DM candidate.   
Of course in all cases, very small couplings would always allow $\chi$ stability, but would definitely be outside the reach of COHERENT.

As already mentioned, in a UV-completed model the possible presence of additional interactions beyond those considered in Eq.~\eqref{eq:Lagr} could be relevant for the production of $\chi$ as the DM in the early Universe. We notice that small couplings of the order of those probed by COHERENT already indicate a possible nonthermal production, e.g. via the freeze-in mechanism.
We conclude by mentioning that, even if stable, light (sub-)MeV DM could affect the predictions for the BBN, thus requiring a careful evaluation of all cosmological constraints.

Finally, going back to the region of parameter space where the DF is unstable, one can investigate whether it decays fast enough to give rise to novel signatures inside the COHERENT detector.
Based on the COHERENT sensitivity reach demonstrated e.g. in Fig.~\ref{fig:proportional-mchi-vs_mv-vector-mediator}, we find that $\chi$ can decay inside the detector only when $m_\chi > m_V$ i.e., via the $\chi \to V \nu_\alpha$ mode. However, if $m_V$ is also larger than twice the electron mass,  $V$ can subsequently decay into a $e^- e^+$ pair via
\begin{equation}
    \Gamma(V \to e^- e^+) =\frac{g_f^2}{12 \pi } m_V\left(1 + 2 \frac{
   m_e^2}{m_V^2}\right)  \sqrt{1-4\frac{m_e^2}{m_V^2}} \, ,
\end{equation}
and possibly lead to a detectable electronic recoil signal within the COHERENT detector.

\section{Conclusions}
\label{sec:concl}
In this paper we have investigated the possible production of a dark fermion $\chi$ at the COHERENT experiment, through neutrinos up-scattering off both the nuclei and the atomic electrons of the detector. We have performed a detailed statistical analysis of two COHERENT datasets: the most recent one, obtained with the CsI detector (2021) and the LAr one (2020). With the idea of going beyond up-scattering through the dipole operator previously studied in the literature, we have focused on two possible mediators: a light vector and a light scalar. We have provided the relevant cross sections for the processes $\nu \mathcal{N} \to \chi \mathcal{N}$ and  $\nu e \to \chi e$, for both vector and scalar mediators.

We have obtained $90 \%$ C.L. exclusion regions on the relevant coupling and mediator/dark fermion mass parameter space. Our results show that including the ES in the CsI analysis improves the bounds on the couplings only in the vector-mediator case. In the scalar-mediator scenario the ES events turn out not to be relevant, nevertheless the most recent CsI data still allow to improve the sensitivity on the couplings, compared to LAr data. Furthermore, in comparison with other experiments and under the assumption of a very light dark fermion, our combined analysis improves the existing bounds for mediators with masses in the range $70~\mathrm{MeV} \lesssim m_V \lesssim 110~\mathrm{MeV}$ and $m_S \gtrsim 40$. Future data from COHERENT will allow to further probe the possible up-scattering production of a dark fermion.

We have finally briefly discussed the possibility for $\chi$ to comprise the dark matter of the Universe. We have found that, when assuming a very light mediator, given the masses and couplings accessible at COHERENT, the dark fermion would not be stable enough to be the dark matter. In other regions of the parameter space, the dark fermion could be stable however a careful discussion of its role as the cosmological dark matter would require a dedicated analysis.

\section*{Acknowledgments}
We thank Sergio Pastor for reading this manuscript and for discussions. This work has been partially supported by the Spanish grants No. PID2020-113775GB-I00 (MCIN/AEI/10.13039/ 501100011033) and CIPROM/2021/054 (Generalitat Valenciana). V.D.R acknowledges financial support by the CIDEXG/2022/20 grant (project ``D'AMAGAT'') funded by Generalitat Valenciana. P.M.C is supported by the grant No. CIACIF/2021/281 also funded by Generalitat Valenciana. D.K.P was supported by the Hellenic Foundation for Research and Innovation (H.F.R.I.) under the “3rd Call for H.F.R.I. Research Projects to support Post-Doctoral Researchers” (Project Number: 7036).

\noindent

\appendix
\section{DETAILS OF THE CROSS SECTION CALCULATION}
\label{sec:appendixA}
In this appendix we provide the necessary steps for obtaining the up-scattering cross section formulae given in Eqs.~(\ref{eq:xsec_DF_V}) and (\ref{eq:xsec_DF_S}). To this purpose, starting from the quark-level Lagrangians in Eq.~\eqref{eq:Lagr} we need to follow two steps: first we translate the quark-level cross sections to the nucleon level and then from nucleons to the nucleus (see below for details).

\subsection{Vector mediator}
For the process $\nu_\alpha (p_1) f(p_2) \to \chi (p_3) f(p_4)$, the corresponding tree-level amplitude at the quark level is
\begin{align}
\label{eq:vector-amplitude-complete}
    i\mathcal{M}_V =& \, \dfrac{i}{t-m_V^2} \, \left[\overline{u}_\chi (p_3, s_3) \gamma_\mu g_{\chi_L} P_L u_\nu (p_1, s_1)\right] \left\{ \overline{u}_f (p_4, s_4) \gamma^\mu \left(g_{f_L} P_L + g_{f_R} P_R\right) u_f (p_2, s_2)\vphantom{\dfrac{m_\chi m_f}{m_V^2}}\right. \nonumber \\
    &+ \left.\dfrac{m_\chi m_f}{m_V^2} \, \overline{u}_f (p_4, s_4) \left[\left(g_{f_L} - g_{f_R}\right)P_L + \left(g_{f_R} - g_{f_L}\right)P_R\right] u_f (p_2, s_2)\right\} ,
\end{align}
where the Dirac equation $\not\hspace{-0.3em}p u(p,s) = m u(p,s)$ has been used together with the fact that only left-handed neutrinos and right-handed antineutrinos exist within the SM, thus $P_R\, u_\nu(p_1, s_1) = 0$. The amplitude has been rewritten in this way to highlight the explicit cancellation of axial terms, under the assumption of chirality-blind couplings. Indeed, in order to simplify the calculations, in what follows  the left- and right-handed components of the charged-fermion couplings are assumed to be equal, i.e., $g_{f_L} = g_{f_R} \equiv g_f$. In addition, quark universal couplings are assumed,  $g_u = g_d = g_f$. Finally we also work under the approximation of massless neutrinos, i.e., $m_\nu \to 0$. With this in mind, the amplitude given in Eq.~\eqref{eq:vector-amplitude-complete} is simplified considerably and eventually reads
\begin{equation}
\label{eq:vector-amplitude-shorted}
    i\mathcal{M}_V = \, \dfrac{i}{t-m_V^2} \, g_{\chi_L} g_f \left[\overline{u}_\chi (p_3, s_3) \gamma_\mu P_L u_\nu (p_1, s_1)\right]\left[ \overline{u}_f (p_4, s_4) \gamma^\mu u_f (p_2, s_2)\right] \, .
\end{equation}
Notice that only the vector component survives, while the axial and pseudoscalar components vanish.

Next, we go from quarks to nucleons. This is achieved by computing the quark-operator matrix elements between nucleon states following the procedure usually adopted for DM direct detection searches and described in~\cite{DelNobile:2021wmp,Cirelli:2013ufw}. For a vector-type interaction, we make the following change
\begin{equation}
    g_f \overline{u}_f \gamma^\mu u_f \rightarrow \sum_{f = u,\,d} g_f \left(c_f^{(p)} \overline{u}_p \gamma^\mu u_p + c_f^{(n)} \overline{u}_n \gamma^\mu u_n \right),
\end{equation}
where the coefficients for the proton and neutron are $c_u^{(p)} = c_d^{(n)} = 2$, $c_d^{(p)} = c_u^{(n)} = 1$.

In order to compute the scattering amplitude for the whole nucleus, an additional step is required.  Let $Z$ be the number of protons inside the nucleus and $N = A - Z$ the number of neutrons, with $A$ being the mass number, then one should proceed by changing
\begin{equation}
\label{eq:nucleons-to-nuclei-vector}
     g_f \left(\overline{u}_p \gamma^\mu u_p + \overline{u}_n \gamma^\mu u_n \right) \rightarrow g_f \left(Z F_{W_p} + N F_{W_n} \right)\overline{u}_\mathcal{N}\gamma^\mu u_\mathcal{N},
\end{equation}
where $F_{W_p}$ and $F_{W_n}$ are the nuclear form factors for protons and neutrons, respectively. Finally, we assume that both form factors are equal~\footnote{This is an accurate assumption for the typical momentum transfer involved at COHERENT.}, $F_{W_p} = F_{W_n} \equiv F_W$, we apply the transformations described above and perform the quark summation to obtain
\begin{equation}
\label{eq:vector-amplitude-3}
    i\mathcal{M}_V = \, \dfrac{i}{t-m_V^2} \, 3 g_{\chi_L} g_f F_W(\qtransfer^2) A \left[\overline{u}_\chi (p_3, s_3) \gamma_\mu P_L u_\nu (p_1, s_1)\right]\left[ \overline{u}_\mathcal{N} (p_4, s_4) \gamma^\mu u_\mathcal{N} (p_2, s_2)\right].
\end{equation}

We define the vector coupling $g_V \equiv \sqrt{g_{\chi_L} g_f}$  to reduce the number of parameters by one. At this point, one has just to perform the spin summation, rewrite the four-momenta in the lab frame and compute the cross section as a function of the nuclear recoil energy to obtain Eq.~\eqref{eq:xsec_DF_V}.

Once we have computed the \cevns~cross section, obtaining the ES one is fairly simple. Since quarks and electrons are fermions, their quantum fields have the same structure, hence we can start with Eq.~\eqref{eq:vector-amplitude-shorted} and take $f=e^-$. Quark universality of $g_f$ is extended to fermion universality, which now includes electrons, i.e., $g_f \equiv g_e = g_u = g_d$. The cross section obtained with this amplitude corresponds to a neutrino up-scattering off a single and isolated electron. However, we need to take into account that electrons are bound inside the atomic nucleus. To obtain Eq.~\eqref{eq:xsec_DF_ES_V}, it is required to weigh the free ES cross section with the effective charge, $Z_{\text{eff}}^\mathcal{A}$, corresponding to an energy deposition $E_\text{er}$. The effective charges for the Cs and I isotopes are given in Appendix~\ref{sec:appendixB}.

\subsection{Scalar mediator}
The procedure followed to obtain the DF-production cross section through a scalar interaction is very similar to the one described above for the case of a vector mediator.  For the process $\nu_\alpha (p_1) f(p_2) \to \chi (p_3) f(p_4)$, the corresponding tree-level amplitude is
\begin{equation}
\label{eq:scalar-amplitude-complete}
    i\mathcal{M}_S = \, \dfrac{i}{t-m_S^2} \left[\overline{u}_\chi (p_3, s_3) g_{\chi_L} P_L u_\nu (p_1, s_1)\right] \left[ \overline{u}_f (p_4, s_4) \left(g_{f_L} P_L + g_{f_R} P_R\right) u_f (p_2, s_2)\right].
\end{equation}

Like before, we assume that the left- and right-handed components of the couplings are the same, and also that fermion universality is satisfied.  Next, we go from quarks to nucleons by computing the relevant quark current in nucleons. For a scalar-type interaction, it proves convenient to perform the following modification~\cite{DelNobile:2021wmp,Cirelli:2013ufw}
\begin{equation}
    g_f \overline{u}_f u_f \rightarrow \sum_{f = u,\,d} g_f \left(\dfrac{m_p}{m_f} f_{T_f}^{(p)} \,\overline{u}_p u_p + \dfrac{m_n}{m_f} f_{T_f}^{(n)} \,\overline{u}_n u_n \right),
\end{equation}
where $m_p$ is the proton mass, $m_n$ is the neutron mass, $m_f$ is the quark mass and $f_{T_f}^{(p)}$ and $f_{T_f}^{(n)}$ express the quark-mass contributions to the nucleon mass. The chosen values are given in Eq.~\eqref{eq:quark-mass-f-values}.

In the last step, we translate from the nucleon to the nuclear level in the same way as done for the vector interaction (see Eq.~\eqref{eq:nucleons-to-nuclei-vector}). After performing the quark summation, we obtain 
\begin{equation}
\label{eq:scalar-amplitude-2}
    i\mathcal{M}_S = \, \dfrac{i}{t - m_S^2} C_S^2 F_W(\qtransfer^2) \left[\overline{u}_\chi (p_3, s_3) P_L u_\nu (p_1, s_1)\right] \left[\overline{u}_\mathcal{N}(p_4, s_4) u_\mathcal{N}(p_2, s_2)\right],
\end{equation}
where $C_S^2$ is the scalar coupling found in Eq.~\eqref{eq:scalar_coupling}. We have defined $g_S \equiv \sqrt{g_{\chi_L} g_f}$ to reduce the number of free parameters. Given this amplitude one can compute the differential cross section in Eq.~\eqref{eq:xsec_DF_S}. To obtain the ES one, we start from the amplitude in Eq.~\eqref{eq:scalar-amplitude-complete} and repeat the procedure explained previously for the case of a vector mediator. The result is given in Eq.~\eqref{eq:xsec_DF_ES_S}.

\section{EFFECTIVE ELECTRON CHARGE FOR Cs AND I}
\label{sec:appendixB}

In this Appendix we provide the effective electron charge for Cs and I (see table~\ref{tab:ZeffCsI}). 

\begin{table}[!htb]
\centering
    \begin{tabular}{@{}cc@{ $< E_{\mathrm{er}} \leq$ }c@{}}
    \toprule
    $\boldsymbol{Z^\mathrm{Cs}_\mathrm{eff}}$ & \multicolumn{2}{c}{$\boldsymbol{E_{\mathrm{er}}~(}$\textbf{keV}$\mathbf{)}$} \\
    \midrule
    55 & \multicolumn{2}{c}{$E_{\mathrm{er}} >$ 35.99} \\[2pt]
    53 & 5.71 & 35.99 \\[2pt]
    51 & 5.36 & 5.71 \\[2pt]
    49 & 5.01 & 5.36 \\[2pt]
    45 & 1.21 & 5.01 \\[2pt]
    43 & 1.07 & 1.21 \\[2pt]
    41 & 1.00 & 1.07 \\[2pt]
    37 & 0.74 & 1.00 \\[2pt]
    33 & 0.73 & 0.74 \\[2pt]
    27 & 0.23 & 0.73 \\[2pt]
    25 & 0.17 & 0.23 \\[2pt]
    23 & 0.16 & 0.17 \\[2pt]
    19 & \multicolumn{2}{c}{$E_{\mathrm{er}} \leq$ 0.16} \\
    \bottomrule
    \end{tabular}
    \hfil
    \begin{tabular}{@{}cc@{ $< E_{\mathrm{er}} \leq$ }c@{}}
    \toprule
    $\boldsymbol{Z^\mathrm{I}_\mathrm{eff}}$ & \multicolumn{2}{c}{$\boldsymbol{E_{\mathrm{er}}~(}$\textbf{keV}$\mathbf{)}$} \\
    \midrule
    53 & \multicolumn{2}{c}{$E_{\mathrm{er}} >$ 33.17} \\[2pt]
    51 & 5.19 & 33.17 \\[2pt]
    49 & 4.86 & 5.19 \\[2pt]
    47 & 4.56 & 4.86 \\[2pt]
    43 & 1.07 & 4.56 \\[2pt]
    41 & 0.93 & 1.07 \\[2pt]
    39 & 0.88 & 0.93 \\[2pt]
    35 & 0.63 & 0.88 \\[2pt]
    31 & 0.62 & 0.63 \\[2pt]
    25 & 0.19 & 0.62 \\[2pt]
    23 & 0.124 & 0.19 \\[2pt]
    21 & 0.123 & 0.124 \\[2pt]
    17 & \multicolumn{2}{c}{$E_{\mathrm{er}} \leq$ 0.123} \\
    \bottomrule
    \end{tabular}
    \caption{\centering{Effective electron charge for Cs and I as a function of the energy deposition $E_\mathrm{er}$}~\cite{Thompson:booklet}.}
    \label{tab:ZeffCsI}
\end{table}

\section{DETECTOR-SPECIFIC QUANTITIES FOR CsI AND LAr}
\label{sec:appendixC}
We convert the true nuclear recoil energy into electron equivalent energy $ E_\mathrm{er}$ through the QF. For the CsI detector this is done in terms of the light yield $\text{LY} = 13.35~\mathrm{PE/keV_{ee}}$, with PE$= \mathrm{LY}\times E_\mathrm{er}$, where PE denotes the number of photoelectrons while the electron-equivalent energy $ E_\mathrm{er}$ is obtained from the scintillation curve, as $E_\mathrm{er}= x_1 E_{\mathrm{nr}}^{\prime} + x_2 E_{\mathrm{nr}}^{\prime 2} +x_3 E_{\mathrm{nr}}^{\prime 3} + x_4 E_{\mathrm{nr}}^{\prime 4}$ ($x_1 = 0.0554628,~x_2 = 4.30681,~x_3 = -111.707,~x_4 = 840.384$)~\cite{COHERENT:2021xmm}. For the LAr detector the QF is given by $\mathrm{QF}(E_\mathrm{nr}^\prime) = 0.246 + 7.8 \times 10^{-4} E_\mathrm{nr}^\prime (\mathrm{keV_{nr}})$~\cite{COHERENT:2020iec}.

The energy-dependent efficiency for the CsI detector is given by 
\begin{equation}
\label{eq:CsI_E_efficiency}
\epsilon_E(x) = \frac{a}{1+e^{-b(x-c)}}+d \, , 
\end{equation}
where $x = \mathrm{PE} + \alpha_7$ and $a = 1.32045$, $b = 0.285979$, $c = 10.8646$, $d = -0.333322$~\cite{COHERENT:2021xmm}. Following Ref.~\cite{DeRomeri:2022twg} we account for the $1\sigma$ uncertainty on the efficiency curve through the parameter $\alpha_7$, see  Eq.~(\ref{eq:Nth_CsI_chi2}), which is allowed to float freely between $[-1,+1]\times$PE. For the LAr detector there is no analytical efficiency function and we utilize the data provided in Ref.~\cite{COHERENT:2020ybo} without introducing any nuisance parameter.

The time-dependent efficiency at the CsI detector is given by~\cite{COHERENT:2021xmm}
\begin{equation}
\label{eq:CsI_T_efficiency}
  \epsilon_T(t_{\mathrm{rec}}) = \begin{cases}
    1, & \text{if}\ t_{\mathrm{rec}}< t^\prime \\
    e^{-k(t_{\mathrm{rec}}-t^\prime)}, & \text{if}~ t_{\mathrm{rec}}\geq t^\prime 
    \end{cases}
\end{equation}
with $t^\prime = 0.52~\mathrm{\mu s}$ and $k = 0.0494\mathrm{/\mu s}$. In this case we allow for a $\pm 250$ ns variation of the timing distribution by introducing the nuisance parameter $\alpha_6$, with no prior  assigned. For the LAr detector it holds $\epsilon_T(t_{\mathrm{rec}})=1$.

Finally, the resolution function $\mathcal{R}(E_{\text{nr}},E'_{\text{nr}})$ for the CsI detector is given by 
\begin{equation}
\mathcal{R}(E_{\text{nr}},E'_{\text{nr}}) =  \frac{(\mathfrak{a} (1 + \mathfrak{b}))^{1 + \mathfrak{b}}}{\Gamma(1 + \mathfrak{b})} \cdot x^\mathfrak{b} \cdot e^{-\mathfrak{a}(1 +\mathfrak{b})x},
\end{equation}
where $x$ is the reconstructed recoil energy expressed in PE units i.e. PE$(E_{\mathrm{nr}})$, while $\mathfrak{a}$ and $\mathfrak{b}$ instead depend on the true quenched energy deposition: $\mathfrak{a} = 0.0749/E_\mathrm{er}(E'_{\text{nr}})$, $\mathfrak{b} = 9.56 \times E_\mathrm{er}(E'_{\text{nr}})$~\cite{COHERENT:2021xmm}.
For the LAr detector instead, the resolution function is approximated by a normalized Gaussian with resolution power
~\cite{COHERENT:2020ybo}
\begin{equation}
    \dfrac{\sigma_{E_\mathrm{er}}}{ E_\mathrm{er}}= \dfrac{0.58}{\sqrt{E_\mathrm{er}(\mathrm{keV_{ee}})}}.
\end{equation}

\bibliographystyle{utphys}
\bibliography{bibliography}  

\end{document}